\documentclass{aa}
\pdfoutput=1
\usepackage{graphicx}
\usepackage{dcolumn}
\usepackage{amsfonts}
\usepackage{amssymb}
\usepackage{bm}
\usepackage{url}
\usepackage{subfigure}
\usepackage{txfonts}
\usepackage{color}
\usepackage{natbib}

\bibpunct{(}{)}{;}{a}{}{,}

\def\aap{A\&A}

\def\apjs{ApJS}
\def\apjl{ApJL}

\begin{document}
\newcommand{\changeR}[1]{\textcolor{red}{#1}}

\newcommand{\TBB}{{{T_{\rm BB}}}}
\newcommand{\TCMB}{{{T_{\rm CMB}}}}
\newcommand{\Te}{{{T_{\rm e}}}}
\newcommand{\Teq}{{{T^{\rm eq}_{\rm e}}}}
\newcommand{\Ti}{{{T_{\rm i}}}}
\newcommand{\nB}{{{n_{\rm B}}}}
\newcommand{\nHe}{{{n_{\rm ^4He}}}}
\newcommand{\nHet}{{{n_{\rm ^3He}}}}
\newcommand{\nHt}{{{n_{\rm { }^3H}}}}
\newcommand{\nHtw}{{{n_{\rm { }^2H}}}}
\newcommand{\nBes}{{{n_{\rm { }^7Be}}}}
\newcommand{\nLis}{{{n_{\rm { }^7Li}}}}
\newcommand{\nLisi}{{{n_{\rm { }^6Li}}}}
\newcommand{\nS}{{{n_{\rm s}}}}
\newcommand{\Teff}{{{T_{\rm eff}}}}

\newcommand{\id}{{{\rm d}}}
\newcommand{\aR}{{{a_{\rm R}}}}
\newcommand{\bR}{{{b_{\rm R}}}}
\newcommand{\neb}{{{n_{\rm eb}}}}
\newcommand{\neql}{{{n_{\rm eq}}}}
\newcommand{\kB}{{{k_{\rm B}}}}
\newcommand{\EB}{{{E_{\rm B}}}}
\newcommand{\zmin}{{{z_{\rm min}}}}
\newcommand{\zmax}{{{z_{\rm max}}}}
\newcommand{\YBEC}{{{Y_{\rm BEC}}}}
\newcommand{\YSZ}{{{Y_{\rm SZ}}}}
\newcommand{\rhob}{{{\rho_{\rm b}}}}
\newcommand{\Ne}{{{n_{\rm e}}}}
\newcommand{\sigT}{{{\sigma_{\rm T}}}}
\newcommand{\me}{{{m_{\rm e}}}}
\newcommand{\nBB}{{{n_{\rm BB}}}}

\newcommand{\kD}{{{{k_{\rm D}}}}}
\newcommand{\KC}{{{{K_{\rm C}}}}}
\newcommand{\KdC}{{{{K_{\rm dC}}}}}
\newcommand{\Kbr}{{{{K_{\rm br}}}}}
\newcommand{\zdC}{{{{z_{\rm dC}}}}}
\newcommand{\zbr}{{{{z_{\rm br}}}}}
\newcommand{\aC}{{{{a_{\rm C}}}}}
\newcommand{\adC}{{{{a_{\rm dC}}}}}
\newcommand{\abr}{{{{a_{\rm br}}}}}
\newcommand{\gdC}{{{{g_{\rm dC}}}}}
\newcommand{\gbr}{{{{g_{\rm br}}}}}
\newcommand{\gff}{{{{g_{\rm ff}}}}}
\newcommand{\xe}{{{{x_{\rm e}}}}}
\newcommand{\alphafs}{{{{\alpha_{\rm fs}}}}}
\newcommand{\YHe}{{{{Y_{\rm He}}}}}
\newcommand{\SE}{{{\dot{{\mathcal{E}}}}}}
\newcommand{\SQ}{{{{{\mathcal{E}}}}}}
\newcommand{\SN}{{\dot{\mathcal{N}}}}
\newcommand{\Sn}{{{\mathcal{N}}}}
\newcommand{\muc}{{{{\mu_{\rm c}}}}}
\newcommand{\xc}{{{{x_{\rm c}}}}}
\newcommand{\xH}{{{{x_{\rm H}}}}}
\newcommand{\mT}{{{{\mathcal{T}}}}}
\newcommand{\Ob}{{{{\Omega_{\rm b}}}}}
\newcommand{\Or}{{{{\Omega_{\rm r}}}}}
\newcommand{\Odm}{{{{\Omega_{\rm dm}}}}}
\newcommand{\mdm}{{{{m_{\rm WIMP}}}}}
\newcommand{\Acmb}{{{{A_{\rm CMB}}}}}
\newcommand{\Ayco}{{{{A_{\rm y/CO}}}}}
\newcommand{\Ad}{{{{A_{\rm dust}}}}}
\newcommand{\Td}{{{{T_{\rm dust}}}}}
\newcommand{\betad}{{{{\beta_{\rm dust}}}}}
\newcommand{\fyco}{{{{f_{\nu}^{\rm y/CO}}}}}
\newcommand{\nudo}{{{{\nu_{0}^{\rm dust}}}}}
\newcommand{\fnud}{{{{f_{\nu}^{\rm dust}}}}}

\title{An alternative validation strategy for the Planck cluster catalogue
  and $y$-distortion maps}
\titlerunning{Vaildiation of Planck clusters and y-maps}
\author{Rishi Khatri\inst{\ref{inst1}{,}\ref{inst2}}
}
\authorrunning{Khatri}
\institute{Max Planck Institut f\"{u}r Astrophysik,
  Karl-Schwarzschild-Str. 1, 85741 Garching, Germany\\
\label{inst1}
\and 
Tata Institute of Fundamental Research, Homi Bhabha Road, Mumbai 400005,
India\\ 
\email{khatri@theory.tifr.res.in}\label{inst2}
}
\abstract
{We present an all-sky map of the $y$-type distortion  calculated from the full
  mission Planck HFI (high frequency instrument) data using the recently proposed approach to component
  separation, which is based on parametric model fitting and model selection. This simple
  model-selection approach enables us to distinguish between carbon monoxide
  (CO) line
  emission and $y$-type distortion, something that is not possible using
  the internal linear combination based methods. We create a mask to cover the regions
  of significant CO emission relying on the information in the $\chi^2$ map
  that was obtained when fitting for the $y$-distortion and CO emission to the
  lowest four HFI channels.  We
  revisit the second Planck cluster catalogue and try to quantify the quality of the
  cluster candidates in an approach that is similar in spirit  to
  \citet{agha2014}. We find that at least $93\%$ of the clusters in
the cosmology sample are free of CO contamination. We also find that $59\%$ of
unconfirmed candidates may have significant contamination from molecular
clouds. We agree with \citet{planckclusters2015} on the worst offenders.
     We suggest an
  alternative validation strategy of measuring and subtracting the CO emission
  from the Planck cluster candidates using radio telescopes, thus improving the
  reliability of the catalogue.  Our CO mask and annotations to the Planck
  cluster catalogue, identifying cluster candidates with possible CO
  contamination, are made publicly available.
}

\keywords{galaxies: clusters - 
          cosmology: large-scale structure of Universe -
          cosmic  background radiation - Methods: data analysis - Cosmology: observations}
\maketitle

\section{Introduction}
The Planck experiment \citep{planckmission}, with its unprecedented
sensitivity and multi-frequency full-sky coverage, has made it possible for
the first time to create a full-sky map of the $y$-type distortion \citep{plancky,planckymap,hs2014} and detect the $y$-type distortion in hundreds of known
and newly discovered clusters
\citep{planckclusters,planckclusters2015}. The $y$-distortion \citep{zs1969}
maps and cluster catalogues  have
been used to constrain cosmological parameters and cluster physics
\citep{planckclusters2,planckclphy1,planckclphy2,planckclphy3,planckclphy4,hs2014,cosmocluster2015,hda2015,rma2015}.
However about $27.2\%$ of the cluster candidates in the second Planck catalogue are
still unconfirmed. The traditional validation strategy for the 
cluster candidates relies on X-ray and optical observations
\citep{planckvalid1,planckvalid2,planckvalid3}.

The Planck HFI channels, 100, 143, 217, and 353 GHz, which are most useful for the $y$-distortion
studies also encompass the strong CO emission lines in three of the
channels. The Galactic CO emission, with its non-trivial, non-monotonic
spectrum,  is therefore expected to be a
non-negligible contaminant in these channels and hence in any
$y$-distortion maps that are produced using these channels. It is more of a
problem for the $y$-type distortion compared to the cosmic microwave
background (CMB) temperature anisotropies because of the weakness of the
$y$-distortion signal. The methods that have been employed so far to
separate out the $y$-type distortion from Planck maps rely on internal
linear combination (ILC) techniques {\citep{milca,nilc2,planckymap,hs2014}}, which give the best
fit estimates  of the  $y$-distortion on the sky, but cannot quantify the
contamination from other components. An advantage of the ILC based
component separation methods is
that we do not need to know the emission spectrum of other components to separate out the component we are interested in. We follow a
different approach, based on the recently developed linearized iterative
least-squares (LIL) \citep{k2014} parametric model fitting and model
selection algorithm. Our method requires a parametric model to be
specified and therefore needs the spectral form of the other components. An
advantage of the parametric model-fitting is that we get a quantitative
estimate of how good the model fits the data in the residuals or $\chi^2$ of the fit,
which can be used to accept or reject a given model of sky emission. We
specifically use this method to select between  CO emission and $y$-type
distortion in every pixel on the sky, assuming that one or the other
component dominates. Our test also gives an indication when both components
may be present on the sky with neither component dominating, signifying
 for example, that a cluster may be present, but the $y$-type distortion
 signal estimate may be
 contaminated by the CO emission.

We apply our method to the Planck HFI data to construct CO emission and
$y$-type distortion maps and use these maps together with the corresponding
$\chi^2$ maps to construct a CO mask that specifies the minimum
recommended area on the sky that should be masked for cosmological and
$y$-distortion studies.  We also revisit the Planck cluster catalogue and
identify candidates which might in fact be molecular clouds or may have
significant contamination from the CO emission from molecular clouds. This
part of the paper is similar in spirit to the more sophisticated analysis
based on neural networks of \citet{agha2014}, but uses a much
simpler test based on the  $\chi^2$ to select between two competing
models and concentrates on differentiating the Galactic molecular clouds from clusters. Our
analysis suggests an alternative strategy of ``negative validation'' using ground-based radio
telescopes to look for the CO emission in candidates that are strongly
indicated to be molecular clouds. In the direction of real clusters, where we identify
significant CO contamination, radio telescopes can be used to measure and
clean the CO emission from these clusters. This strategy may greatly help
improve the reliability of the
$y$-distortion signal in real clusters and of the cluster catalogue as a
whole.

\section{Construction of full sky map of $y$-type distortion from Planck
  data}

Planck experiment has nine frequency channels covering the frequency range
$30~{\rm GHz}$ to $857~{\rm GHz}$. The
angular resolutions of the lowest three LFI (low frequency instrument)
channels  at $30~{\rm GHz},~ 44~{\rm GHz},~ 70~{\rm GHz}$, are $32',27',13'$,
respectively \citep{lfibeam}. In addition, they have much higher noise
compared to the HFI (high frequency instrument) channels and are therefore
are not suitable for looking for weak and spatially concentrated signals such
as $y$-type distortion. We therefore only use  HFI channels in our
study. The highest two HFI channels at 545 and 857 GHz are dominated by
dust. We  use a simple grey body spectrum with fixed temperature but
varying spectral index to model dust. This is expected to be a good
approximation in the Rayleigh-Jeans region where the spectral shape is not
sensitive to the exact temperature but not accurate enough at 545 GHz and
higher frequencies as we get closer to 
the peak of the grey body  We therefore use the four  frequencies channels at
$100,143,217$, and $353~$GHz to construct a $y$-distortion map and use $545$
GHz channel only to define a dust mask, i.e. a mask covering the regions of
high dust emission above a certain threshold to select different fractions
of the sky.

\subsection{Contamination from $y$-type distortion in CO maps and vice
  versa}  The frequency channels in the Planck experiment are rather broad
and encompass many molecular lines, the most prominent of which are the
${}^{12}{\rm CO}$
emission lines $J=1\longrightarrow 0$, $J=2\longrightarrow
1$,$J=3\longrightarrow 2$ which contribute to 100 GHz, 217 GHz, and 353 GHz
channels \citep{planckco}. We  neglect the contributions from
${}^{13}{\rm CO}$ or other isotopologues. We would expect the CO
  emission from wide redshift bins to contribute to a given frequency
  channel and the same lines from different redshifts may contribute to
  different frequency channels.  We
  use the 545 GHz map to mask strong point sources which should also mask
  most nearby strong sources of CO emission. 
However, except from the nearby galaxies,
  the CO emission contribution from external galaxies
   remains unresolved by Planck and would be part of the diffuse CO
  background which remains in our maps. 

We  want to  mask the regions of high CO
emission. The CO emission has a signal which changes in amplitude
between different frequency channels in a non-monotonic way. Even though this variation is
different from that of the $y$-type distortion, it is enough to present a
serious contaminant as there are not enough frequency channels in Planck to
simultaneously separate the CMB, $y$-type distortion, dust and CO emission.
 This is particularly important because the molecular clouds can
present a morphology very similar to the galaxy clusters and there are
regions on the sky with rather low dust emission but still non-negligible
CO emission. 

In principle, each Planck channel has  many
detectors with slightly different frequency response and it is possible to
utilize the difference in transmission of lines in different detectors of
the same channel to
separate out the CO line emission. Planck collaboration uses this
technique and the resulting maps are labeled as Type 1 maps using {the MILCA
component separation algorithm \citep{milca,planckco}.} These maps are however very noisy and not very useful
except within a few degrees of the galactic plane. The most complete CO line survey
 is  by  \citet{dame2001} who combined 37 individual surveys together
 to  create a composite map of the galactic CO emission and the velocity integrated maps
 can be downloaded from \url{http://lambda.gsfc.nasa.gov/product/foreground/dameco_map.cfm}. Unfortunately this map is also
 confined to mostly latitudes $|b|\lesssim 32^{\circ}$ and  does not
 extend to high latitudes which are of most interest for cosmology. There
 are many smaller  surveys covering the galaxy away from the galactic plane
 in the northern as well as the southern galactic hemispheres \citep{magnani1985,hartmann1998,magnani2000,wilson2005} which have
 revealed the existence of molecular clouds at galactic latitudes up to
 $|b|=55^{\circ}$. With Planck of course we can detect the molecular clouds
 over the full sky. The difficulty, as we will see below, lies in
 separating the CO emission from the $y$-type distortion at high galactic
 latitudes and this is where our algorithm, to be discussed below, is
 efficient and most useful.

\begin{figure*}
\resizebox{12 cm}{!}{\includegraphics{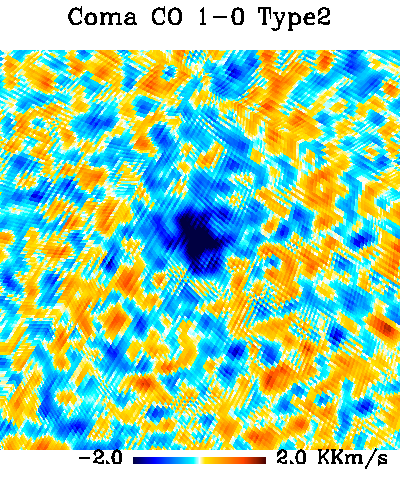}\hspace{10
  pt}\includegraphics{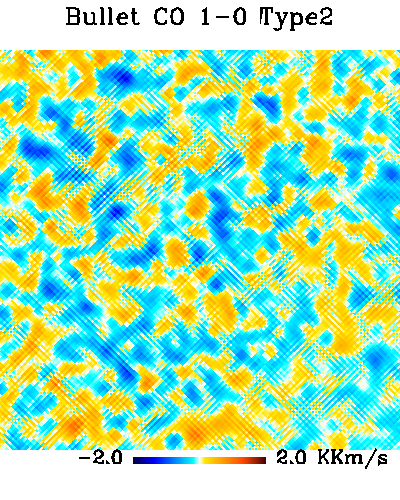}\hspace{10
  pt}\includegraphics{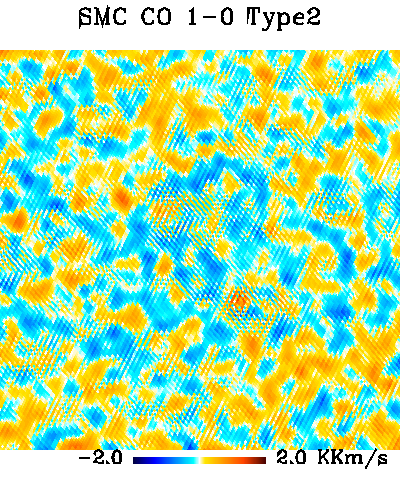}}\\
\sidecaption
\resizebox{12 cm}{!}{\includegraphics{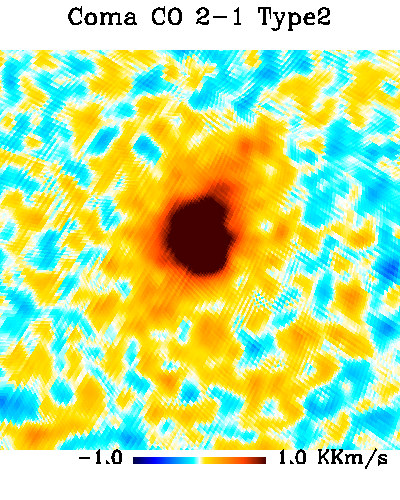}\hspace{10
  pt}\includegraphics{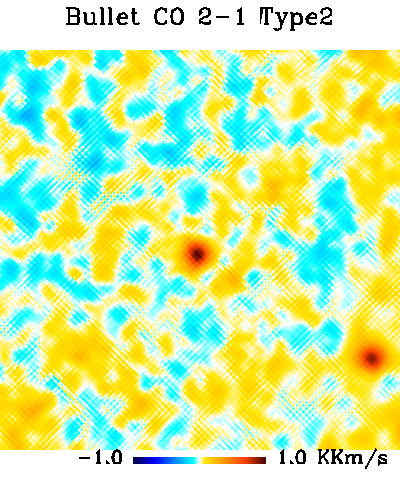}\hspace{10
  pt}\includegraphics{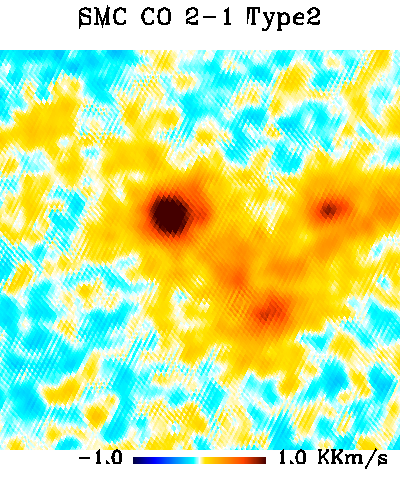}}
\caption{\label{Fig:planckco} Coma and Bullet clusters and SMC in Planck
  released CO Type 2 maps. Clusters show up as fake molecular clouds in $J=2\longrightarrow
1$ maps while  strong clusters show up as negative sources in $J=1\longrightarrow
0$ maps. A 5 degree by 5 degree region around each source is shown.}
\end{figure*}

 The second approach used by the Planck team is to use full channel maps and fit a
 parametric model to separate out the CO component and these are labeled
 Type 2 and Type 3 maps {using the Commander-Ruler  and Ruler
\citep{eriksen2008,planckfg} algorithms.}  Type 2 maps fit separately for the two lowest CO
 transitions while Type 3 maps assume a constant line ratio between the
 lines and fit for the common amplitude. The $y$-type distortion shows up as a negative
 source in Type 2  $J=1\longrightarrow 0$ maps since it contributes to the
 100 GHz channel and as a positive source in $J=2\longrightarrow
1$ maps as this transition corresponds to 217 GHz channel where the
$y$-distortion signal integrated over the band is  positive. In principle
therefore it is possible to use $J=1\longrightarrow 0$ line to distinguish
between the negative $y$-distortion sources and the positive genuine molecular
clouds. This map is however too noisy and misses weak CO sources, for
example 
the small Magellanic clouds (SMC). A mask based on  $J=2\longrightarrow
1$  maps  will also mask all the clusters. To
 illustrate this we show the region around  Coma and Bullet clusters and
 SMC in the publicly released
 Type 2 CO  maps of Planck  in Fig. \ref{Fig:planckco}. Type 3 maps are
 similar to the Type 2  $J=2\longrightarrow
1$  maps. {There is of course negligible CO emission from the Coma and
Bullet clusters and what we see in the CO maps is the $y$-distortion signal
falsely identified as CO signal by the numerical fit to data. Similarly we
expect that the CO emission would be falsely identified as $y$-distortion
signal in the $y$-maps.}

{We  derive the CO maps using the constant line ratios  similar to the Planck type-3
maps which would then have similar contamination from clusters. We then follow a  robust approach based on model selection to
distinguish between the CO line emission and the $y$-type distortion.}

\subsection{Linearized iterative least-squares (LIL) parameter fitting}
We want to fit the following parametric model to the four Planck frequency
channels with frequencies $100,143,217$, and $353~$GHz at each pixel
\begin{align}
s_{\nu}(p)&=\Acmb + \fyco \Ayco(p) \nonumber\\
&+ \Ad(p)\fnud\frac{1}{\exp\left(\frac{h\nu}{\kB \Td}\right)-1}\left(\frac{\nu}{\nudo}\right)^{\betad(p)},
\end{align}
where $A_i$ are the amplitudes of the corresponding components, $p$ is the
pixel index in the map,  $\betad$ is the spectral index for the dust
spectrum and $\nudo=353~{\rm GHz}$. The factor of $\fnud$ takes into account the colour correction for
the dust spectrum \citep{planckhfi} assuming a constant temperature,
$\Td=18.0{\rm K}$. The factor of $\fyco$ is the spectrum of the $y$-type distortion or the CO line
emission, where we fit for either the CO line emission or the $y$-type
distortion at a time. 
\begin{figure}
\resizebox{\hsize}{!}{\includegraphics{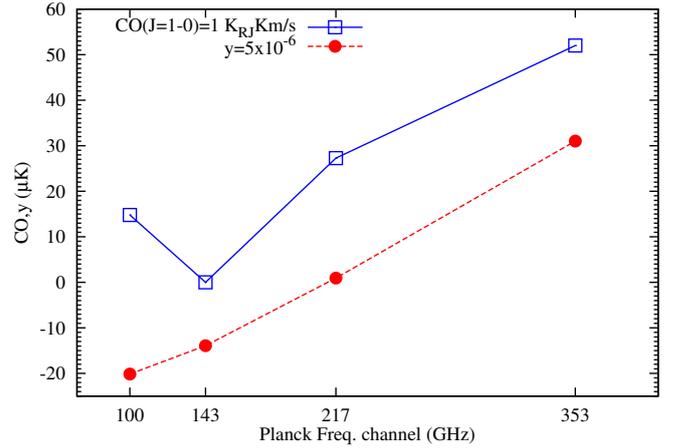}}
\caption{\label{Fig:ycospec} CO and $y$ spectrum as seen by different
  Planck channels after integrating over the detector response and
  conversion to thermodynamic CMB temperature units.}
\end{figure}

{Following \citep{planckco}, we assume constant line
ratios for the CO line contribution to the Planck HFI channels of
1:0.595:0.297  for $(J=1\longrightarrow
0):(J=2\longrightarrow
1):(J=3\longrightarrow
2)$.  The $y$-type spectrum at frequency $\nu$ is given by \citep{zs1969} 
\begin{align}
\Delta I_{\nu}=\frac{2 h\nu^3}{c^2}\frac{x
  e^x}{(e^x-1)^2}\left[x\left(\frac{e^x+1}{e^x-1}\right)-4\right],
\end{align}
where $x=h\nu/(\kB \TCMB)$, $h$ is the Planck constant, $\kB$ is the
Boltzmann constant, $\TCMB=2.725~{\rm K}$ is the CMB temperature and $c$ is
the speed of light. The factor $\fnud$ includes conversion from the Rayleigh-Jeans $({\rm K}_{\rm
  RJ})$
temperature to thermodynamic CMB temperature units ${\rm K}_{\rm CMB}$ for dust
and $\fyco$ includes the conversion from ${\rm K}_{\rm RJ}~{\rm km/s}$ or dimensionless $y$ amplitude  for CO emission
and $y$-type distortion respectively for different frequency channels
\citep{planckhfi} and are obtained by integrating the spectrum over the
frequency response of the detectors. The CO spectrum and $y$-distortion
spectrum as seen by Planck are compared in Fig. \ref{Fig:ycospec}. An
important difference between the two spectra is the rise in the CO spectrum
at 100 GHz compared to the dip in the $y$-type spectrum. They are of course
sufficiently similar that if we just try to extract one component by a
linear combination of different channel maps, the two components will leak
into each other. If only a CO component is present but we fit for a $y$
component, we will expect a non-zero answer since the two components are not
orthogonal. However if instead of doing a linear combination, we fit a
parametric model, then these two spectra are sufficiently different that
even though we would get a non-zero best fit answer for the wrong component,
the fit would be much worse than expected from the available degrees of
freedom, i.e. the residuals between the data and the model ($\chi^2$)
would be very high. This is the key that enables us to separate the CO
and the $y$-type components.}

\begin{figure*}
\resizebox{\hsize}{!}{\includegraphics{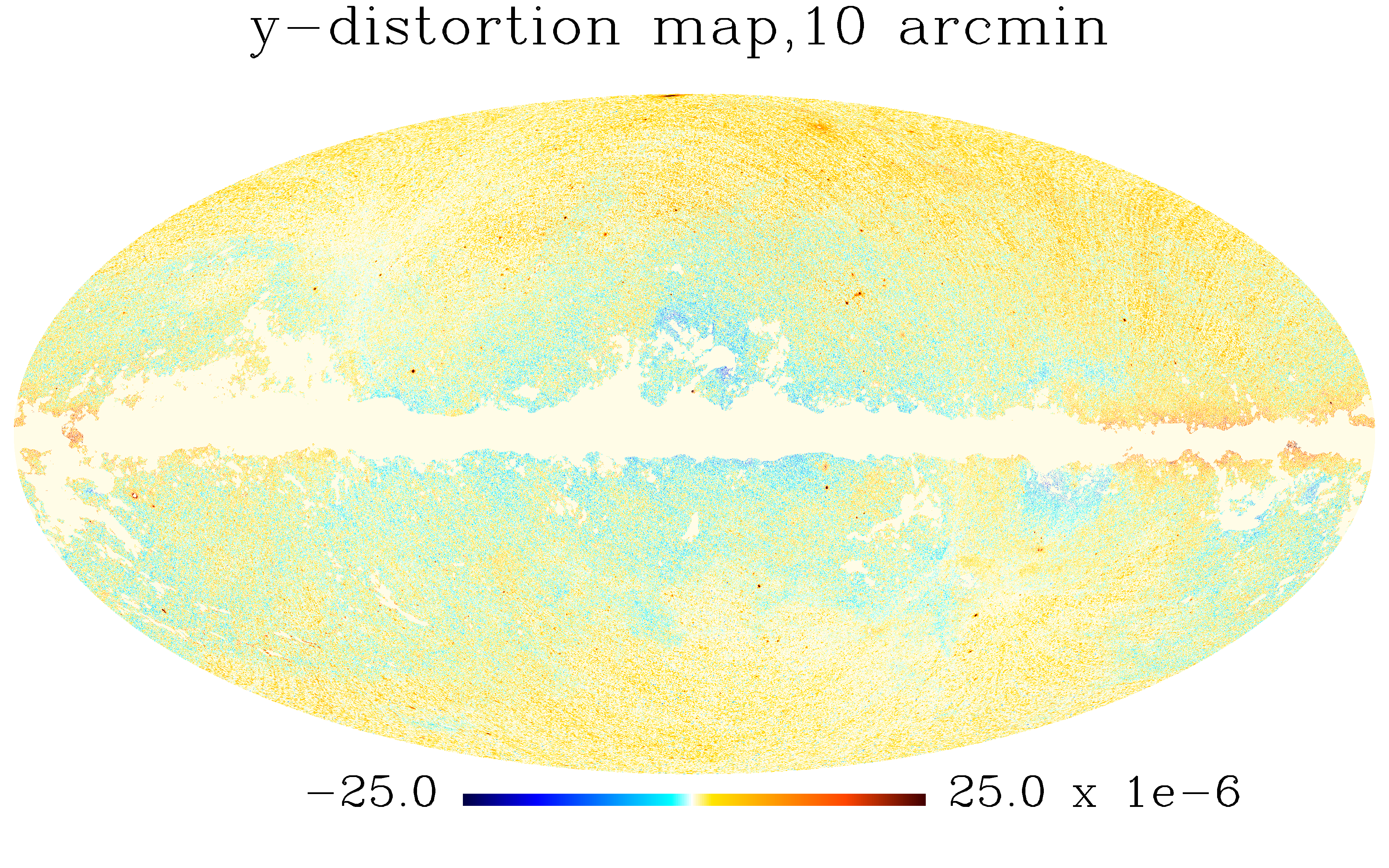}}
\caption{\label{Fig:y}Map of $y$-type distortion at $10'$ resolution
  constructed from the lowest four Planck HFI channels. Approximately
  $14.2\%$ of the area is masked. The prominent clusters can clearly be
  identified. Typical contamination from the foregrounds and noise is of
  order $\sim 10^{-6}$. The ring like systematic features come from the
  Planck scanning strategy.}
\end{figure*}

We use the recently developed LIL algorithm to fit for four parameters
(three amplitudes and one dust spectral index) to four frequency channels at each
pixel. We refer the reader to \citep{k2014} for the details of the
parameter fitting algorithm. Here we just note a few important features of
the algorithm. Since we are fitting four parameters to the four data points, it
would appear that the degrees of freedom are zero. However in our algorithm
the non-linear
parameter is constrained to lie between values of
$2<\betad<3$ \citep{fds1999,gold2011,planckfg,planckdust}  and is not free to
vary freely towards unphysical values where the true global minimum of the
least squares problem may in fact lie. In particular in the regions of low dust contamination we do
not expect to find the global or even local minimum of the $\chi^2$ in the $\betad$ direction,
and the value chosen for the $\betad$ is just the best value within the
constraints which often turns out to be one of the boundaries. Therefore
the effective degree of freedom is usually close to one over most of the
sky. Also the parameters are not arbitrary but belong to  fixed models. So
even while fitting four data points with four parameters, the $\chi^2$
would in general be $\gg 1$ if the wrong model is being fit to the
data. We rely on this feature to choose the correct
model. Finally we want to rebeam all frequency channels to a common
beam/angular resolution so that they are correctly weighted while parameter
fitting. We use the channel beam window functions provided by the Planck
collaboration as the  beam profile of the respective channel maps\citep{hfibeam}.
We then rebeam all the maps to a common $10'$ FWHM Gaussian beam, which is close to the
resolution of the lowest HFI frequency channel of 100 GHz and use the
half-ring maps to estimate the noise in the rebeamed maps.

For both the $y$-type distortion and the CO emission we do nested model selection
while fitting for the parameters as follows: we fit both a three parameter CMB+dust
model and a four parameter CMB+dust+y/CO model. If an additional component
corresponding to y or CO is present, we should get a big improvement in
$\chi^2$ when we fit with the four parameters. The difference in $\chi^2$
for the three and four parameter models is again expected to be distributed as
$\chi^2$ distribution for one degree of freedom \citep{kendall}. If the
y/CO components are absent we do not expect a big improvement in
$\chi^2$. We can therefore set a threshold in $\Delta \chi^2$ improvement
that we get when adding an extra component for accepting that component. If
the improvement is smaller than the threshold we set the y or CO component
to zero. For the CO component additionally we also put the additional constraint that it
should be greater than zero. If during the fitting procedure the CO
component falls below zero, we fix it to zero for the next few steps. A
threshold of $\Delta \chi^2=1.6,2.7,3.8$ would correspond to the $20\%,
10\%,5\%$ probabilities respectively that we accept the y/CO component as
present when it is in fact absent in a particular pixel. We  use
$\Delta \chi^2=3.8$ in the present paper. We have tested our results with
different values of $\Delta \chi^2$, and for the construction of the CO
mask and validation of the cluster catalogue, our results are not sensitive
to the exact value of $\Delta \chi^2$. 

\begin{figure*}
\resizebox{\hsize}{!}{\includegraphics{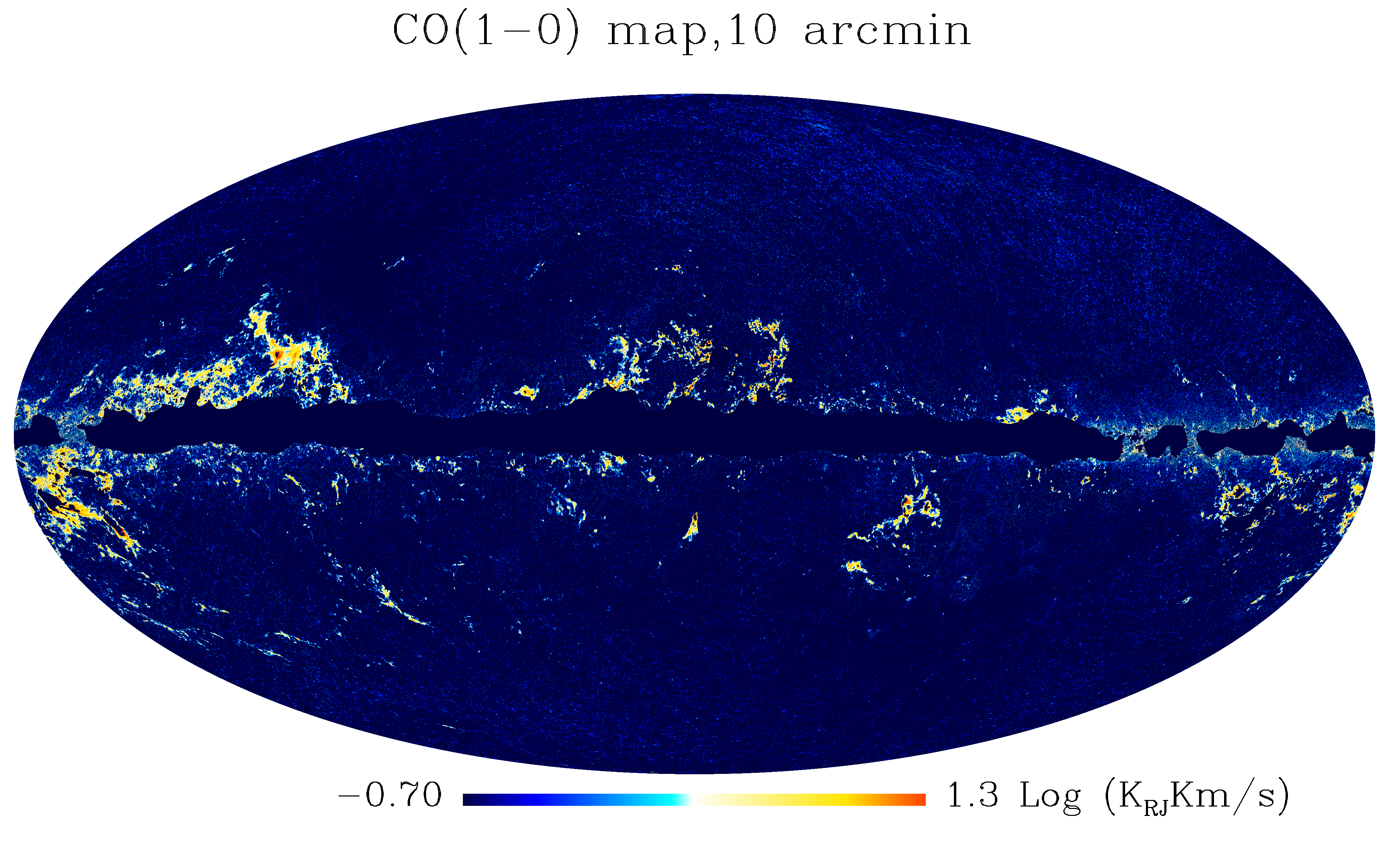}}
\caption{\label{Fig:co}Map of CO emission  at $10'$ resolution
  constructed from the lowest four Planck HFI channels. Approximately
  $10.2\%$ of the area with extremely high dust emission and clusters are
  masked.}
\end{figure*}

The full sky $y$-distortion map computed for the Planck full mission data
 for $\Delta \chi^2=3.8$ is shown in Fig. \ref{Fig:y} with the $14\%$ of the area  masked ($y=0$ inside the masked area). We discuss the
details of the mask computation in the next section. {The
 official Planck full sky $y$-distortion maps\citep{planckymap} computed
 using two algorithms based on the internal linear combination (ILC)
 technique, modified ILC algorithm (MILCA) \cite{milca} and Needlet ILC
 (NILC) \cite{nilc,nilc2} are  publicly
 available and we compare these with our map in section \ref{sec:comp}. We
  also  do an indirect comparison using the publicly available second Planck
 cluster catalogue. In particular, for the average $y$-type
 contribution from the Planck detected clusters we agree giving
 for the clean cluster sample (see below) an average $\langle y \rangle \approx 4\times 10^{-8}$ \citep{ks2015}}.

We also show in Fig. \ref{Fig:co} our CO emission map. We have masked
extremely high dust emission regions as well as the regions where our algorithm
prefers a $y$-type distortion over the CO emission, i.e the clusters. The algorithm used to
create this $y$+dust mask is similar to the one  used for the CO mask
described below.

\section{Model selection between CO and $y$-distortion, construction of
  CO mask and validation on real sky}
\begin{figure*}
\resizebox{12 cm}{!}{\includegraphics{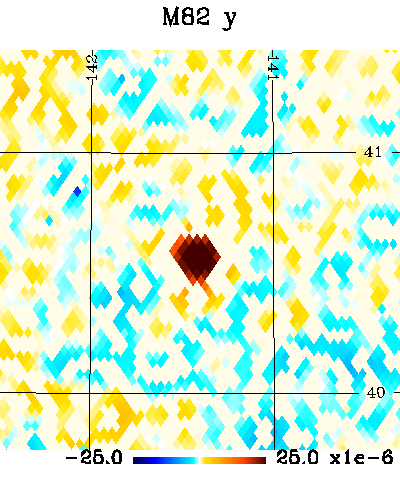}\hspace{10
  pt}\includegraphics{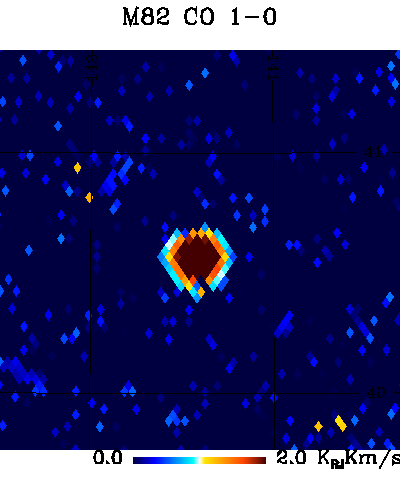}\hspace{10
  pt}\includegraphics{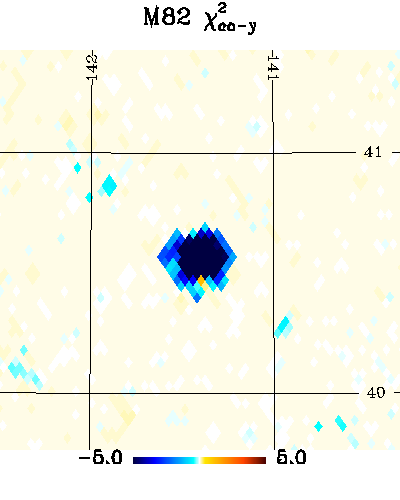}}\\
\resizebox{12 cm}{!}{\includegraphics{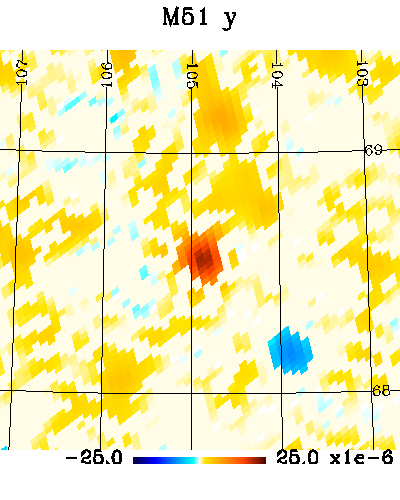}\hspace{10
  pt}\includegraphics{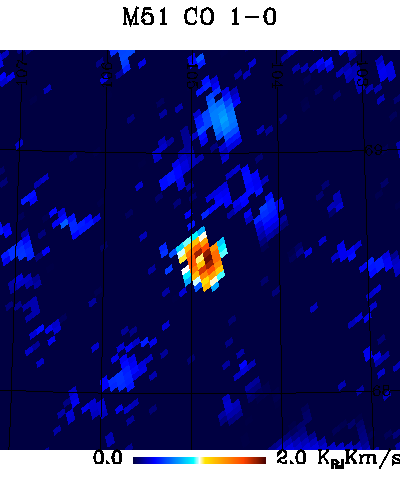}\hspace{10
  pt}\includegraphics{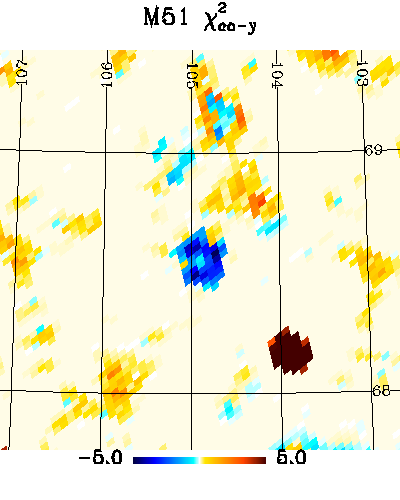}}\\
\resizebox{12 cm}{!}{\includegraphics{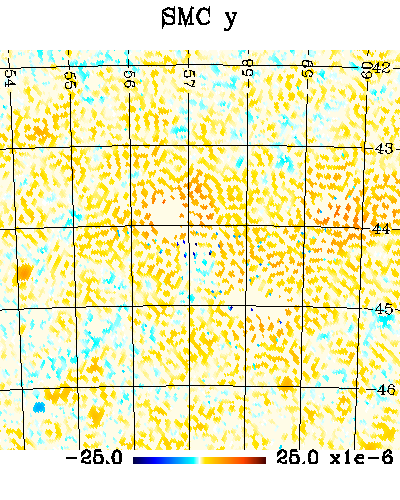}\hspace{10
  pt}\includegraphics{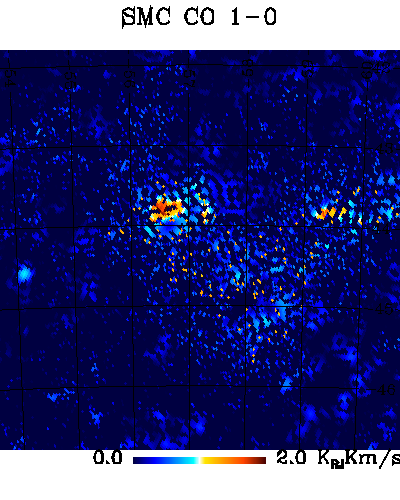}\hspace{10
  pt}\includegraphics{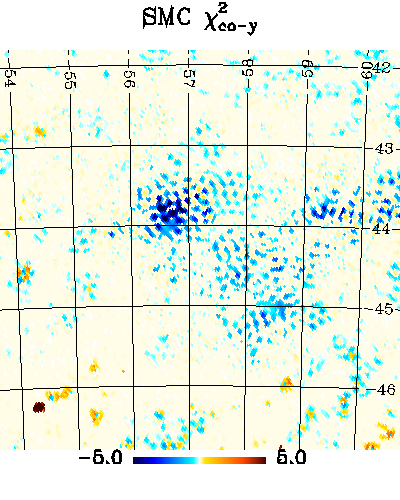}}\\
\sidecaption
\resizebox{12 cm}{!}{\includegraphics{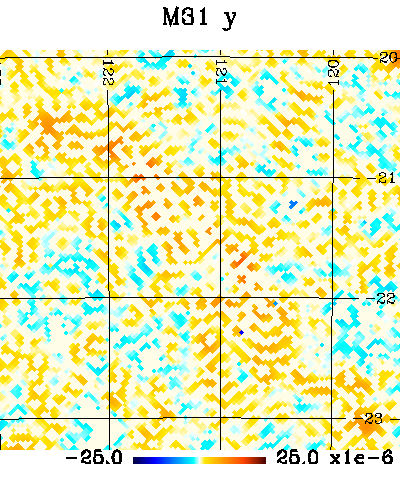}\hspace{10
  pt}\includegraphics{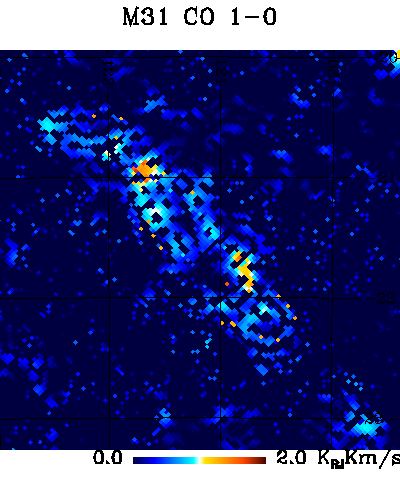}\hspace{10
  pt}\includegraphics{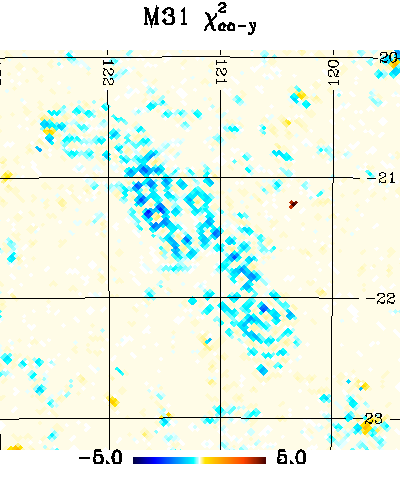}}
\caption{\label{Fig:planckgal} Some external galaxies with different
  strengths and morphology in CO emission in our $y$-distortion and
  CO maps and difference in $\chi^2$ between the two model fits. A negative
  value implies that CO model is favored over the $y$-distortion. A
  negative $y$-distortion source near M51 is a radio point
  source. Galactic coordinates are shown.}
\end{figure*}

{The validation of our algorithm with the FFP6 (full focal plane) \citep{ffp62013} simulations was already
performed in \citet{k2014} for the CO signal where we also compared our CO
maps on the real sky with those released by the Planck collaboration and
found good agreement. It was shown there that our
algorithm recovers quite well the morphology of the CO emission.} In this section we take a different
approach of validating our algorithm as well as the model selection idea on
the real sky using the known extragalactic sources of CO emission and
clusters of galaxies. This is very important for us since the FFP6 simulations
are based on the Dame et al. map \citep{dame2001} which as mentioned above
does not cover high galactic latitudes where we want to do cosmological
analysis. Note that this is the second level of model selection that we
would perform now, in addition to the nested model selection already mentioned in
the previous section. We now want to compare two models, CMB+dust+y and
CMB+dust+CO, each with the same number of parameters. It is possible to use
an information criterion to select between the models by looking at
the difference $\chi^2_{\rm CO}-\chi^2_{\rm y}$. Different information
criterions differ mostly in how they penalize the new parameters \citep{kendall}. Since the number of
parameters and degrees of freedom for our two models are same, the
different information criterions such as Akaike information criterion \citep{akaike1974},
Bayesian information criterion \citep{schwarz1978} and Hannan-Quinn
 information criterion \citep{hq1979} are equivalent for us. 
We  use  the CO and $y$ distortion maps created with the
nested  model
selection threshold of $\Delta \chi^2=3.8$ as discussed in the previous
section.

\begin{figure*}
\resizebox{\hsize}{!}{\includegraphics{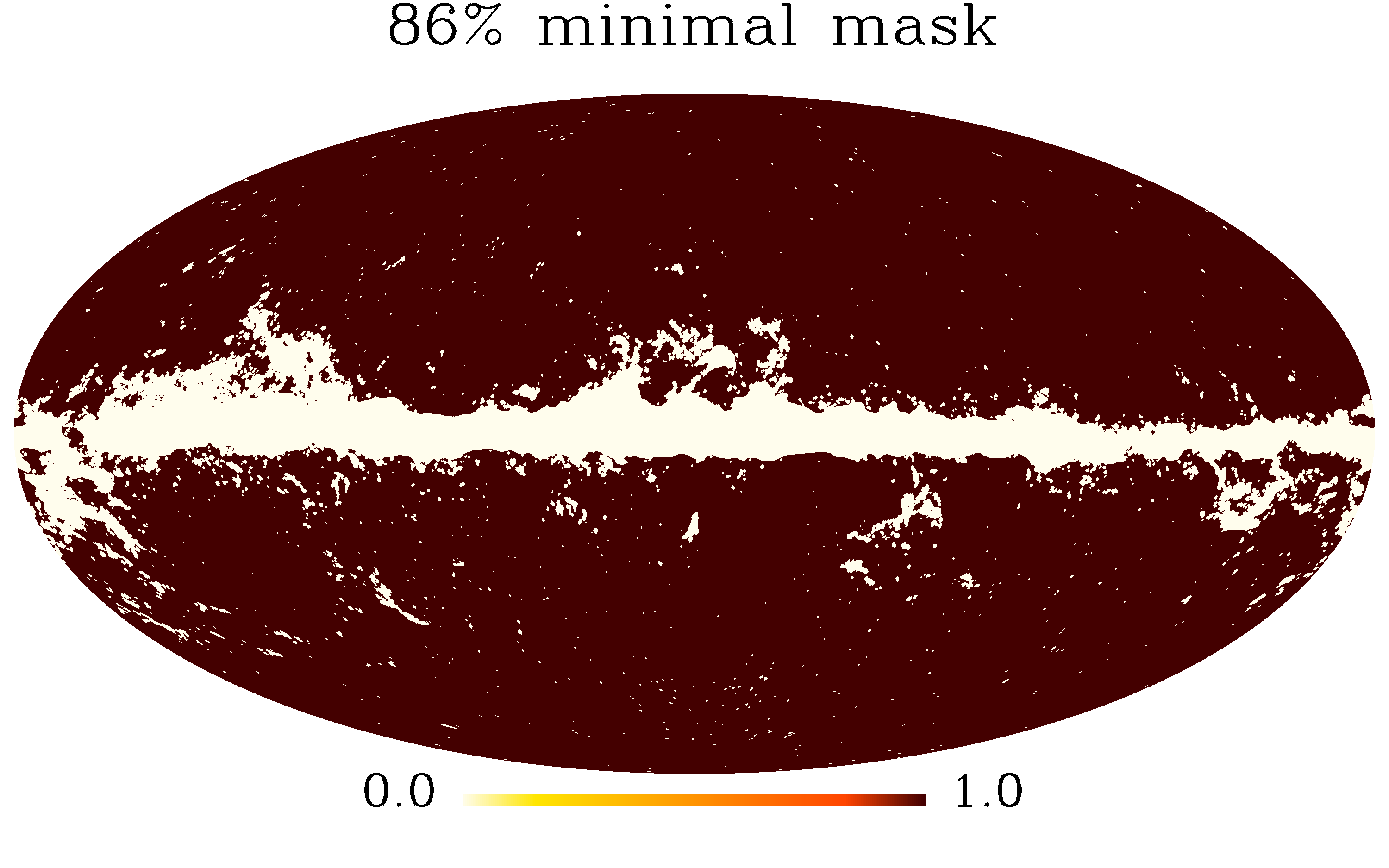}}
\caption{\label{Fig:mask86} Minimal mask covering $14.16\%$ of the
  sky and used for Fig. \ref{Fig:y}.}
\end{figure*}

{We show in Fig. \ref{Fig:planckgal} our CO map, $y$ map and $\Delta
\chi^2_{\rm CO-y}\equiv \chi^2_{\rm CO}-\chi^2_{\rm y}$ map several
external galaxies. The M82 (Cigar galaxy) and M51 (Whirlpool galaxy) are
unresolved strong sources of CO emission and are clearly identified as such
in the $\chi^2_{\rm CO-y}$ map which takes large negative values. SMC is a resolved source with varying
surface brightness across it. The morphology of our signal is in agreement
with the that obtained from the dedicated CO $(J=1\longrightarrow
0)$ line observations
done with the NANTEN millimeter-wave telescope \citep{nanten}. Even a weak
 but large  
source as Andromeda (M31) is clearly identified as a CO source and the
outer spiral arm which has the strongest CO emission can clearly be
distinguished in the CO and $\chi^2$ maps. This ring like morphology of the
CO emission in Andromeda agrees with the CO $(J=1\longrightarrow
0)$ observations of \citep{nngu2006} with the IRAM
30-m telescope. We have also detected the Antennae galaxies and M101 or Pinwheel
galaxies in our CO maps as weak (as far as signal in Planck is concerned) sources of CO emission. We of course also
detect the large
Magellanic cloud (LMC) as a strong source of CO emission with morphology that
again agrees with the dedicated observations from the  Magellanic Mopra
assessment (MAGMA) survey \citep{magma}. The LMC  can be clearly identified
in our CO mask below (Fig: \ref{Fig:mask86}). {We note that all
  these galaxies show significant $y$-distortion signal which is of course
  the CO emission contamination.} Our $\chi^2$ test
identifies the signal as CO emission in all the cases.} Thus our algorithm
can cleanly identify and 
mask out the sources of CO emission. 

\begin{figure*}
\resizebox{12 cm}{!}{\includegraphics{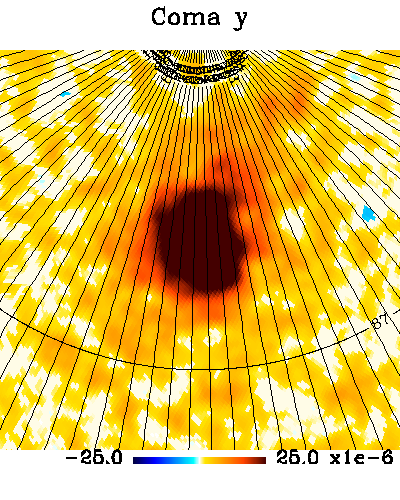}\hspace{10
  pt}\includegraphics{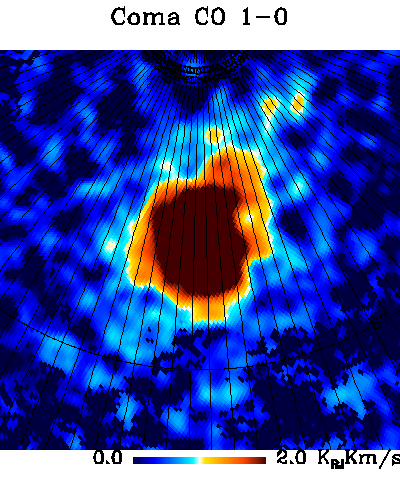}\hspace{10
  pt}\includegraphics{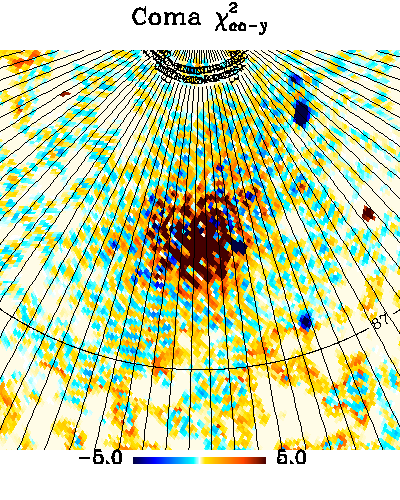}}\\
\resizebox{12 cm}{!}{\includegraphics{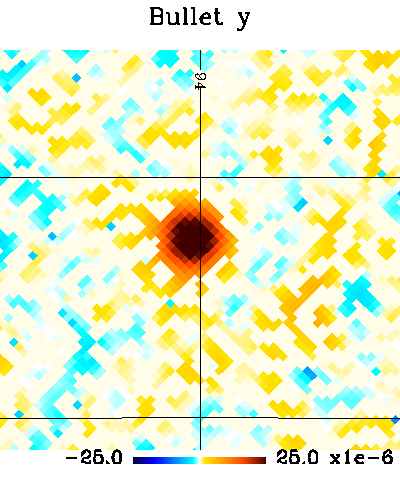}\hspace{10
  pt}\includegraphics{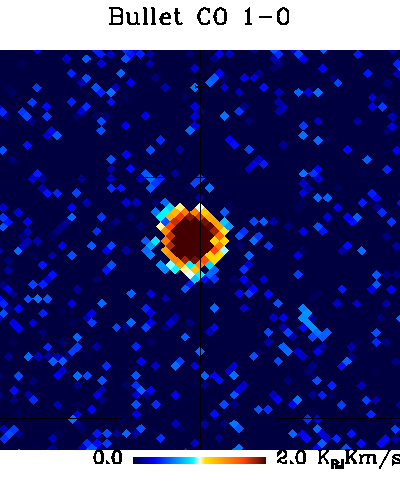}\hspace{10
  pt}\includegraphics{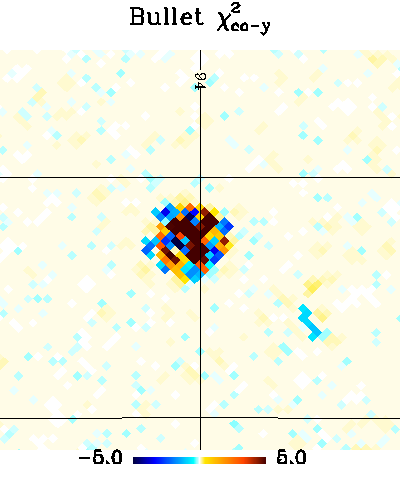}}\\
\resizebox{12 cm}{!}{\includegraphics{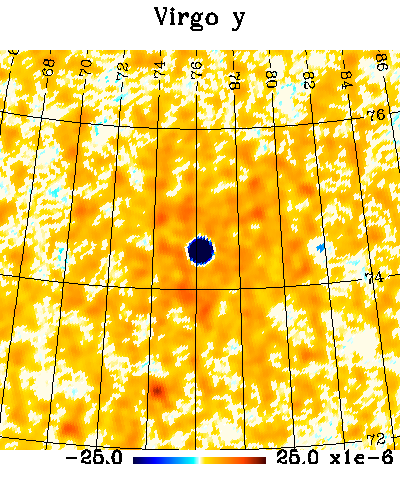}\hspace{10
  pt}\includegraphics{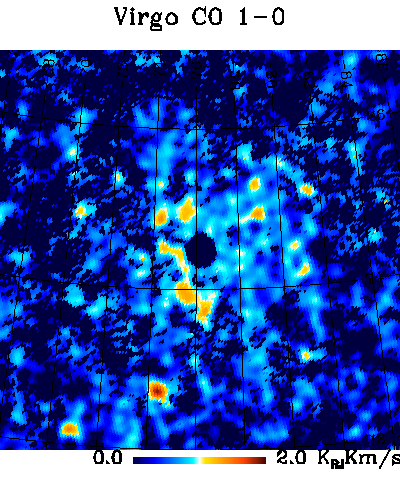}\hspace{10
  pt}\includegraphics{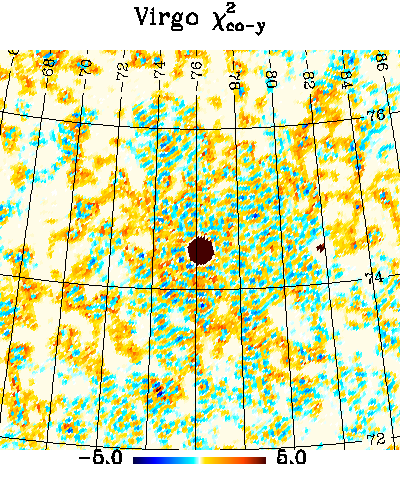}}\\
\sidecaption
\resizebox{12 cm}{!}{\includegraphics{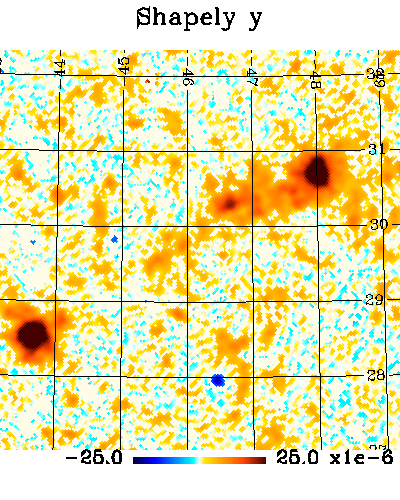}\hspace{10
  pt}\includegraphics{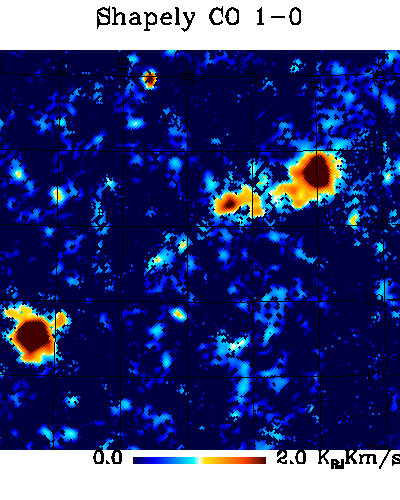}\hspace{10
  pt}\includegraphics{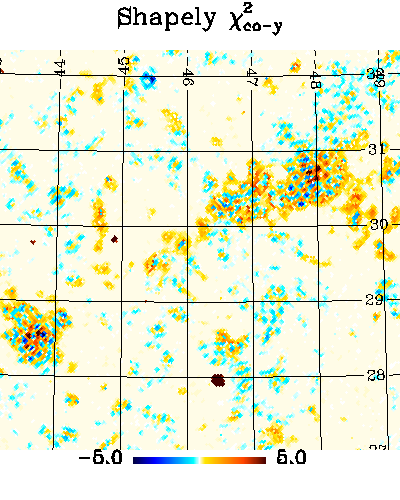}}
\caption{\label{Fig:plancksz} Same as figure \ref{Fig:planckgal} but for the
well known galaxy clusters. The $\chi^2_{\rm CO-y}$ map at the position
has when averaged over many pixels large positive values favoring the
$y$-distortion model over the CO emission. The prominent clusters in the
last panels are A3571, A3562, and A3558 in the Shapley supercluster.}
\end{figure*}

We  show in Fig. \ref{Fig:plancksz} the maps for some of the known
clusters. In this case the $\chi^2_{\rm CO-y}$ takes on average large
positive values at the position of clusters
signifying that the $y$-distortion model is preferred  over the CO emission
model.  There is a bright radio source in the centre of Virgo cluster which
shows up as negative $y$-distortion source. We discuss the handling of
these sources also below when we create the CO mask. 

The comparison of figures \ref{Fig:planckgal} and \ref{Fig:plancksz} shows
that we can cleanly distinguish between CO emission and $y$-type distortion
using the $\chi^2_{\rm CO-y}$ map and create a CO mask for $y$-distortion
studies and vice versa. We use the following algorithm to create the mask.
\begin{enumerate}
\item We first mask all pixels for which $\chi^2_{\rm CO-y}\le -1$ and also
  those pixels for which $\chi^2$ for the $y$-map is $\chi^2_y>10.0$
  i.e. the $y$-distortion model was a very bad fit even without comparing
  with the CO map.
\item We then unmask all the pixels which were just masked as a result of
  statistical fluctuation by filling holes (masked regions) in the mask
  with size smaller than 50 pixels (in the  HEALPix\citep{healpix}
  nside=2048 resolution maps) i.e. a region to be masked must have at least
  50 pixels. This makes sure that only genuine CO regions
  are masked. 
\item We also want to mask point sources which may be smaller than 50
  pixels in size. For this we repeat the first step with higher threshold
  $\chi^2_{\rm CO-y}\le -2$. In addition we also mask pixels with large
  negative values of $y$-type distortion $y<-50$, and pixels with negative
  values and high $\chi^2$, $y<0 ~\& ~\chi^2_{\rm y}> 1.6$. We then fill
  holes in the mask with a lower threshold of 12 pixels.
\item Finally we increase the mask by three pixels around the point sources and
  by five pixels everywhere else.
\end{enumerate}

The above threshold were chosen so that we mask the known CO sources such
as those shown
in Fig. \ref{Fig:planckgal} while preserving even weak (in $y$-distortion) known clusters. There
is a trade-off here: By being aggressive we can mask even the weak CO
sources at the expense of deleting some weak clusters. By being
conservative we can keep even the weak $y$-distortion signals at the
expense of some contamination from the weak CO sources. Our thresholds are
conservative and  mask sources such as M51, M82, and SMC but do
not mask Andromeda, the latter gives very weak contamination to the
$y$-distortion map as shown in Fig. \ref{Fig:planckgal}. We augment our mask
slightly to mask regions with very high dust contamination using the 545
GHz channel. Our final mask covers $14.16\%$ of sky leaving $85.84\%$ of
the sky unmasked. This is our minimal recommended mask that should be
included in all studies of $y$-distortion in Planck data and is shown in
Fig. \ref{Fig:mask86}. In practice we would further augment this mask using
545 GHz channel to get cleaner and cleaner parts of the sky to test the
robustness of our results to contamination by dust. Our mask is made
publicly available at \url{http://theory.tifr.res.in/~khatri/szresults/}.

\begin{figure}
\resizebox{\hsize}{!}{\includegraphics{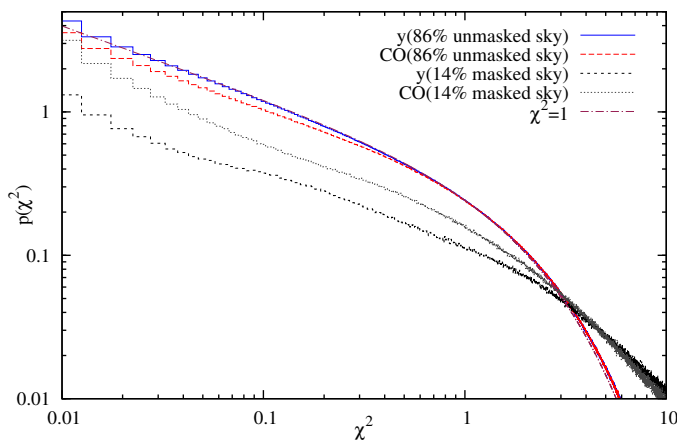}}
\caption{\label{fig:chisq}  $\chi^2$ distribution for the parameter fits
by LIL for the $y$-distortion model and CO model compared with the $\chi^2$
distribution for one degree of freedom.}
\end{figure}

We show in Fig. \ref{fig:chisq} the $\chi^2$ distribution for pixels when
fitting a $y$-distortion signal and a CO signal inside and outside our
minimal $86\%$ sky fraction mask (with the nested model selection turned
off, $\Delta \chi^2=0$). As expected the $\chi^2$
distribution outside the mask is close to the $\chi^2$ distribution for one
degree of freedom, also shown in the plot, signifying that our model is a
good fit. The $\chi^2$ distribution is also
slightly better (lower $\chi^2$ values) for a $y$ distortion fit
outside the mask compared to the CO spectrum. Inside the mask, the $\chi^2$
is slightly smaller for the model with CO, as expected. Note that for our
model selection approach, only the difference in $\chi^2$ is
important. Therefore, even though from the $\chi^2$ plot we see that inside
the mask our model is not a good fit because the foregrounds become too
complicated for our simple model, which ignores in particular the
synchrotron component and fixes the CO spectrum, it still tells us that a $y$-distortion
signal is a worse fit compared to the CO. 

\section{Comparison with the MILCA/NILC $y$-maps}\label{sec:comp}
Planck collaboration has released two $y$-distortion maps computed using
the internal linear combination (ILC) methods. The MILCA algorithm uses the HFI
from 100 GHz to 545 GHz, and the 857 GHz channel on large angular
scales. The NILC algorithm in addition also uses the LFI data on large
angular scales, $\ell < 300$ \cite{planckymap}. Both algorithms filter the
maps with harmonic space window functions, apply ILC, and recombine the
resulting $y$ maps in $\ell$ windows to form the final $y$ map. The main
difference between the two algorithms is the bandpass window functions which for
NILC correspond to the Needlet decomposition. The main advantage of
performing an ILC is that we do not need to specify a model for the
contamination. In particular the MILCA and NILC  both remove a
combination of the dust and CO foregrounds. This is the main difference
from LIL where we explicitly remove the dust component but not the CO
component. The LIL maps are therefore expected to have slightly higher
level of CO contamination, which should not be a problem since the CO
signal presents non-negligible contamination in only a small fraction of
the sky which we mask. Note that since we explicitly take out the dust
contamination modeled by a grey body spectrum, any other source of
contamination with similar spectrum, such as the cosmic infrared background
(CIB), is automatically fitted out by LIL. In this section we  use the
LIL $y$-map with $\Delta \chi^2=0$, i.e. with nested model selection turned
off and $y$-type component included in all pixels.

Even though NILC and MILCA use higher number of channels, it is not enough
to remove all CO and dust contamination. This is because some of the additional
information coming from the 545 GHz channel is offset by the additional
complexity of modeling the dust emission over a wider frequency range. 
\begin{figure*}
\resizebox{12 cm}{!}{\includegraphics{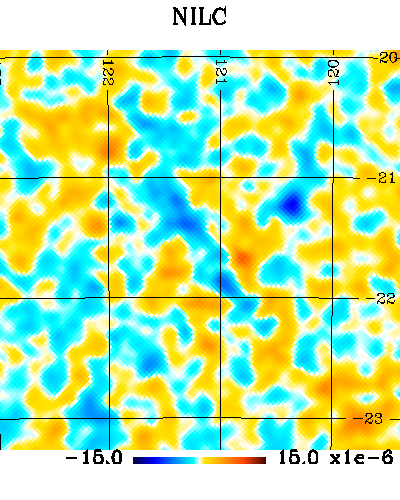}\hspace{10
  pt}\includegraphics{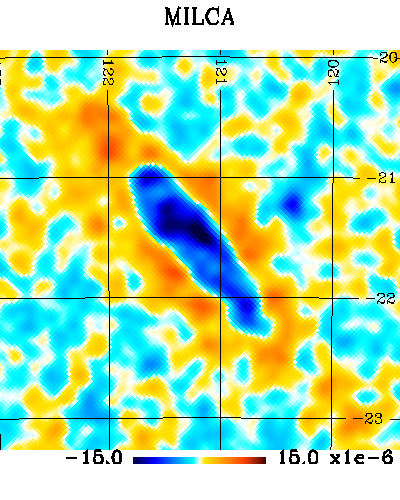}\hspace{10
  pt}\includegraphics{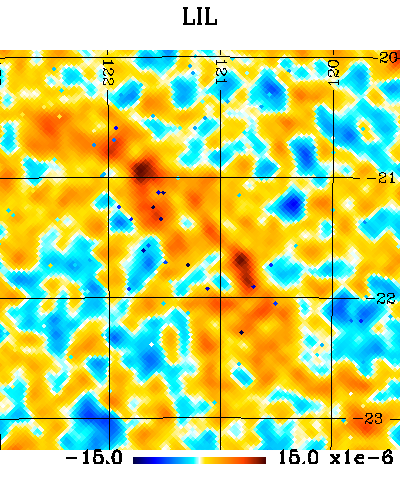}}\\
\resizebox{12 cm}{!}{\includegraphics{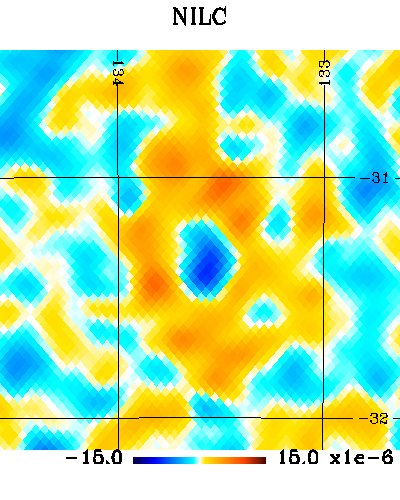}\hspace{10
  pt}\includegraphics{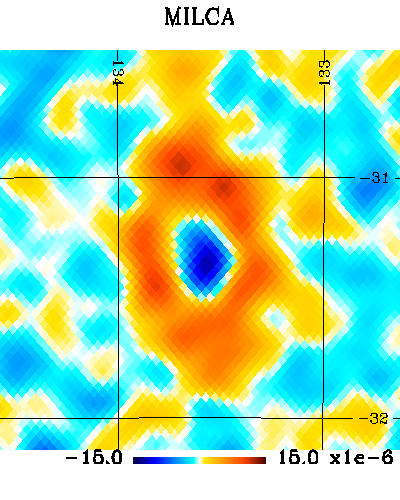}\hspace{10
  pt}\includegraphics{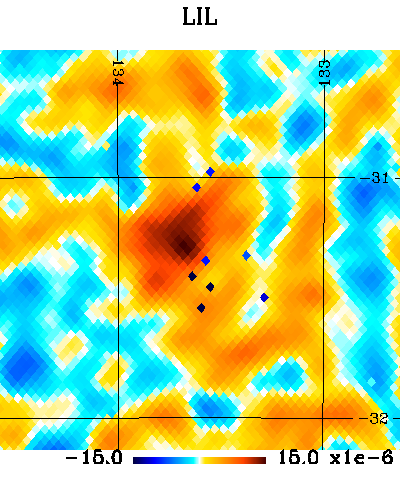}}\\
\resizebox{12 cm}{!}{\includegraphics{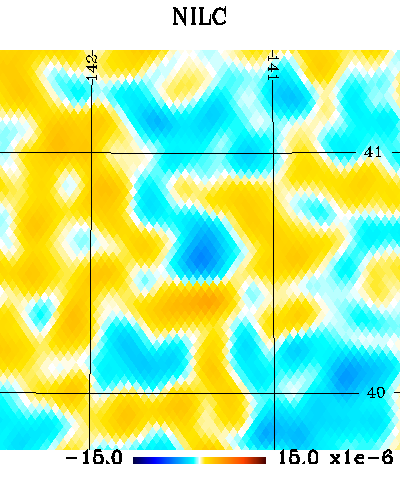}\hspace{10
  pt}\includegraphics{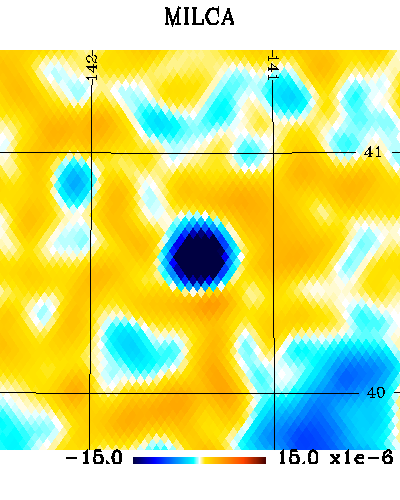}\hspace{10
  pt}\includegraphics{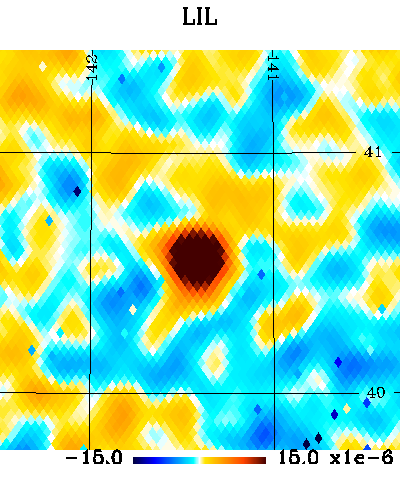}}\\
\sidecaption
\resizebox{12 cm}{!}{\includegraphics{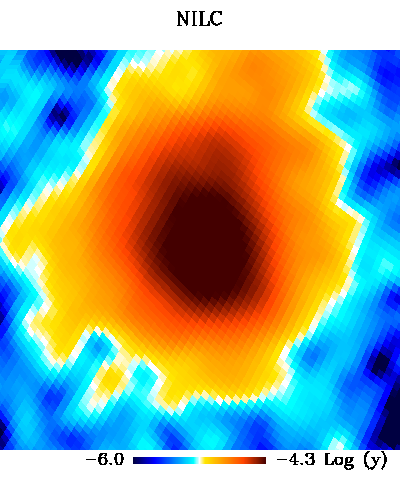}\hspace{10
  pt}\includegraphics{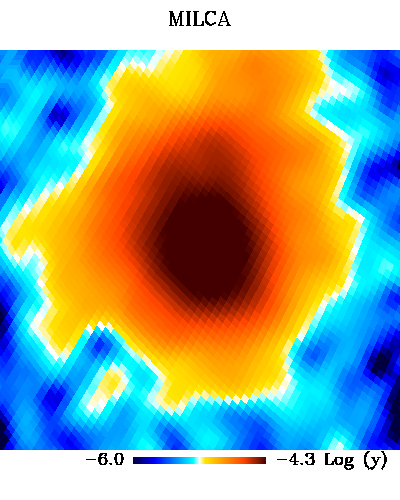}\hspace{10
  pt}\includegraphics{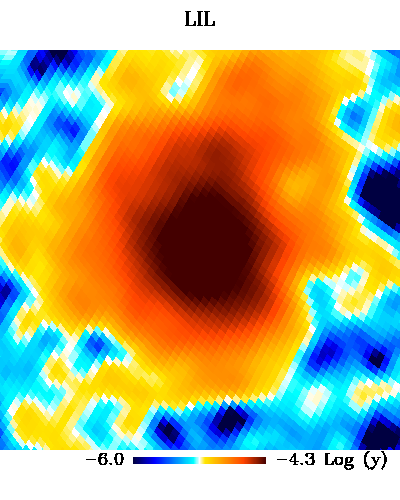}}
\caption{\label{Fig:comp} Some external galaxies with different
  strengths and morphology in CO and dust emission are compared for
  different component separation methods. From top to bottom the galaxies
  shown are M31, M33, and M82. The last row shows the Coma cluster.}
\end{figure*}
We compare the three nearby galaxies, M31, M33, and M82 in
Fig. \ref{Fig:comp}, in NILC, MILCA, and LIL $y$-maps. These external
galaxies, with known dust and CO emission morphologies, are  good
realistic test cases to  study the extent of residual contamination in the
$y$-maps. The contamination in the LIL map is correlated with the CO
emission from these galaxies, as expected, and shows up in bright clumps in
the spiral arms. For MILCA and NILC maps, the contamination has
contribution from both residual dust and CO and is therefore more diffuse
and is negative as well as positive. The contamination is slightly less in
the NILC maps compared to the MILCA maps. We also show Coma cluster in the
three $y$-maps in the last panel of Fig.  \ref{Fig:comp} and the cluster
$y$ signal agrees between the two maps. The comparison of these external galaxies therefore
reveals the complementarity of the parameter fitting LIL approach vs ILC
approach. 

We should clarify here that the main goal of our new method is not to
produce a better $y$-map or to exactly reproduce the results of the Planck
collaboration. Our main goal is to identify the regions of the sky with
significant CO contamination. This is a worthwhile goal, since the CO
contamination is a rare signal on the sky, just like the $y$-distortion
signal. Since it occupies a small fraction of the sky area, it is possible to use targeted radio observations from ground to measure and
remove the CO lines from the Planck maps. On these CO-free maps we can then
apply the ILC/LIL methods to produce much cleaner and accurate $y$-distortion
maps, since the ILC/LIL will now have less components to contend with. As we
saw in the previous section, the $\chi^2$ based model selection, a standard
statistics method \cite{kendall}, performs quite well in this regard.

\begin{figure}
\resizebox{\hsize}{!}{\includegraphics{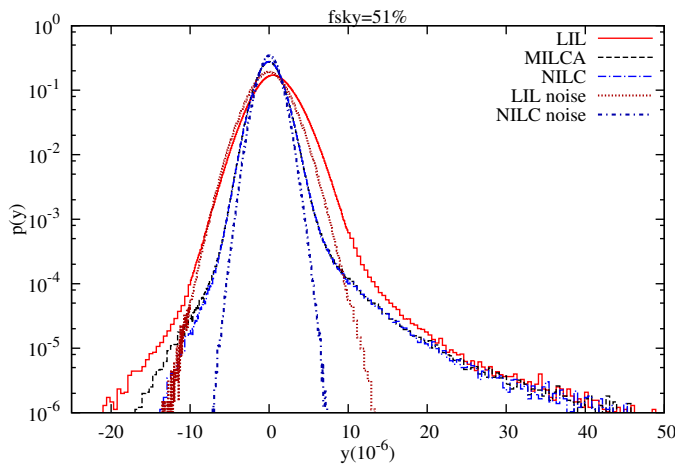}}
\caption{\label{Fig:1d51}  1-d PDF for NILC, MILCA, and LIL $y$-maps for
  the same mask with $51\%$ sky.}
\end{figure}

\begin{figure}
\resizebox{\hsize}{!}{\includegraphics{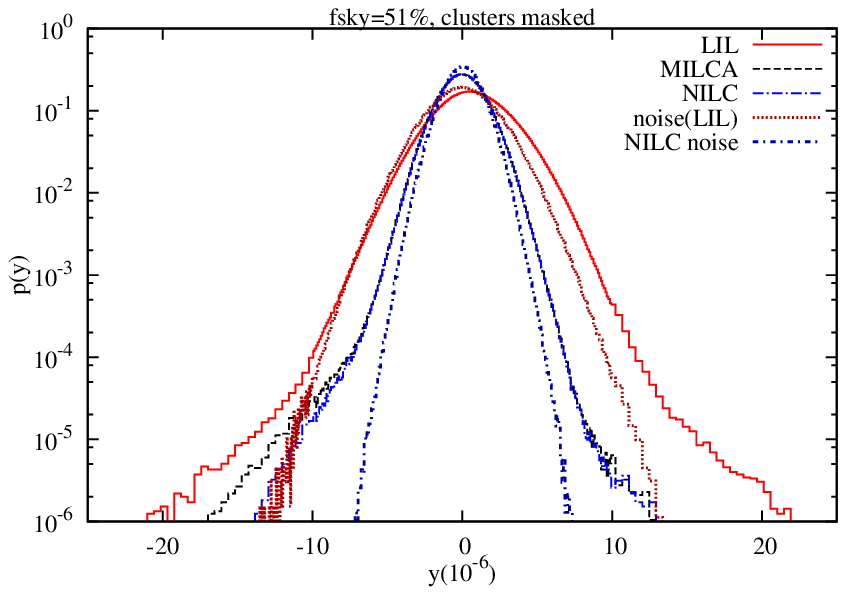}}
\caption{\label{Fig:1d51cl}  1-d PDF for NILC, MILCA, and LIL $y$-maps for
  the same mask with $51\%$ sky and in addition with all the clusters from the second Planck
  cluster catalogue masked.}
\end{figure}

We show in Fig. \ref{Fig:1d51} the comparison of the 1-d probability distribution
functions (PDF), $p(y)$, for the three maps for $51\%$ of the sky utilizing
the mask described in the previous section and augmenting it with a
galactic dust mask constructed from the Planck 545 GHz map.\footnote{All
  masks used in this section are publicly available at \url{http://theory.tifr.res.in/~khatri/szresults/}.} 
The tails at high values of $y$, contributed mostly by the already detected clusters
of galaxies, agree for all maps. 
We also show the noise PDFs for LIL and NILC maps calculated from the half
ring half difference (HRHD) maps. The MILCA noise is similar to NILC and
both have smaller noise compared to the LIL maps. An important difference
between NILC/MILCA and LIL is seen near the peak. The NILC/MILCA PDF is
symmetric around the peak a result of explicitly removing the monopole in
their component separation algorithms. In the LIL method we do not subtract
the monopole and the PDF is skewed at small $y$ as predicted \cite{rs2003}. This is apparent when we
compare the $y$ PDF with the noise PDF, the later being a symmetric
Gaussian. This skewness is expected if we have unresolved $y$ signal
contributed by weak sources such as filaments and groups of galaxies. This
is apparent in Fig. \ref{Fig:1d51cl} where we masked  $0.75^{\circ}$ radius
regions around all clusters and
cluster candidates from the second Planck cluster catalogue \cite{planckclusters2015},  with slightly
bigger masks of $1.5^{\circ}$ for sources with S/N>30 and $4.5^{\circ}$
for Perseus and Virgo.

\begin{figure}
\resizebox{\hsize}{!}{\includegraphics{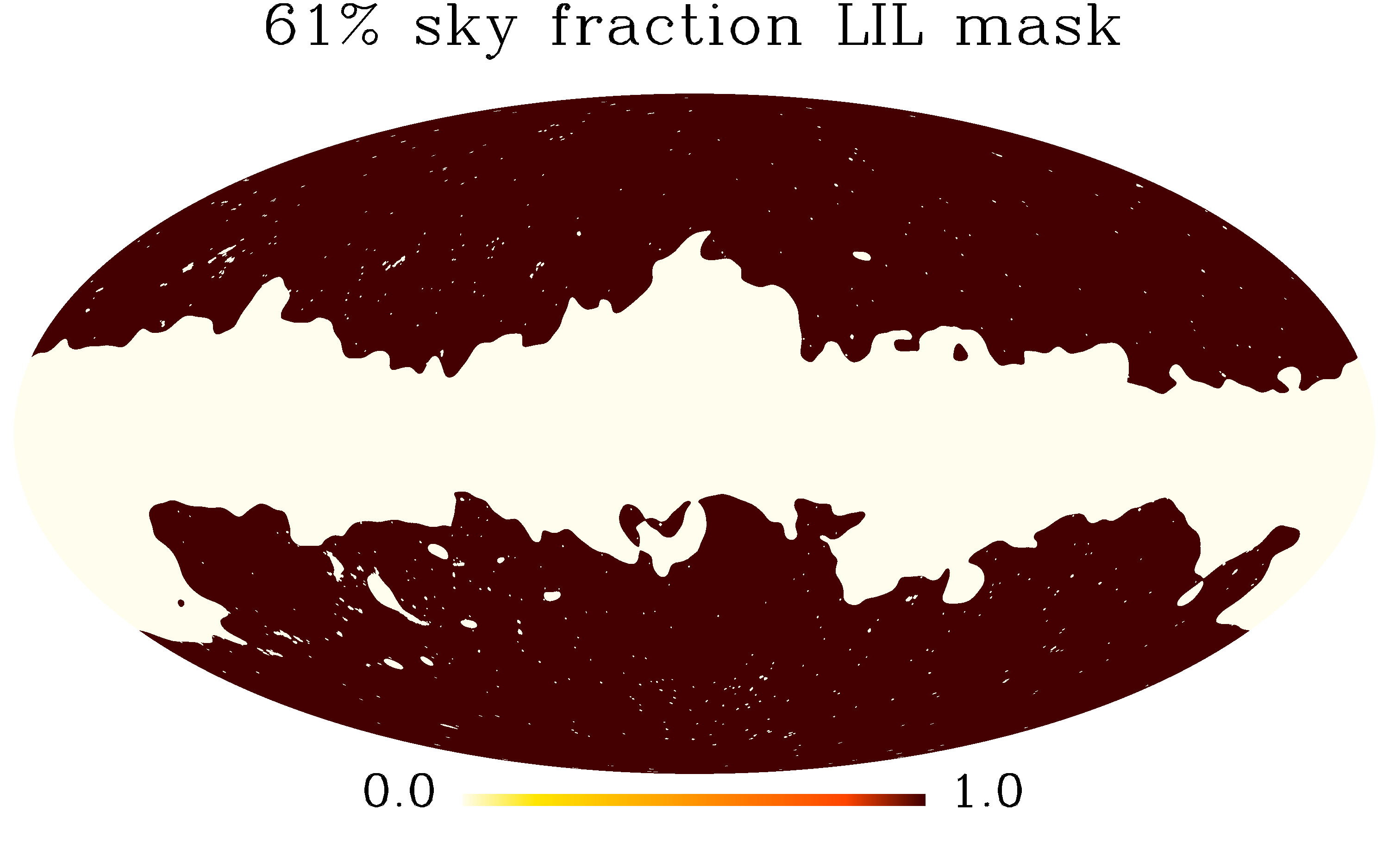}}
\caption{\label{fig:masklil} Our mask with $61\%$ sky fraction calculated
  using the model selection based approach described in the previous section.}
\end{figure}

\begin{figure}
\resizebox{\hsize}{!}{\includegraphics{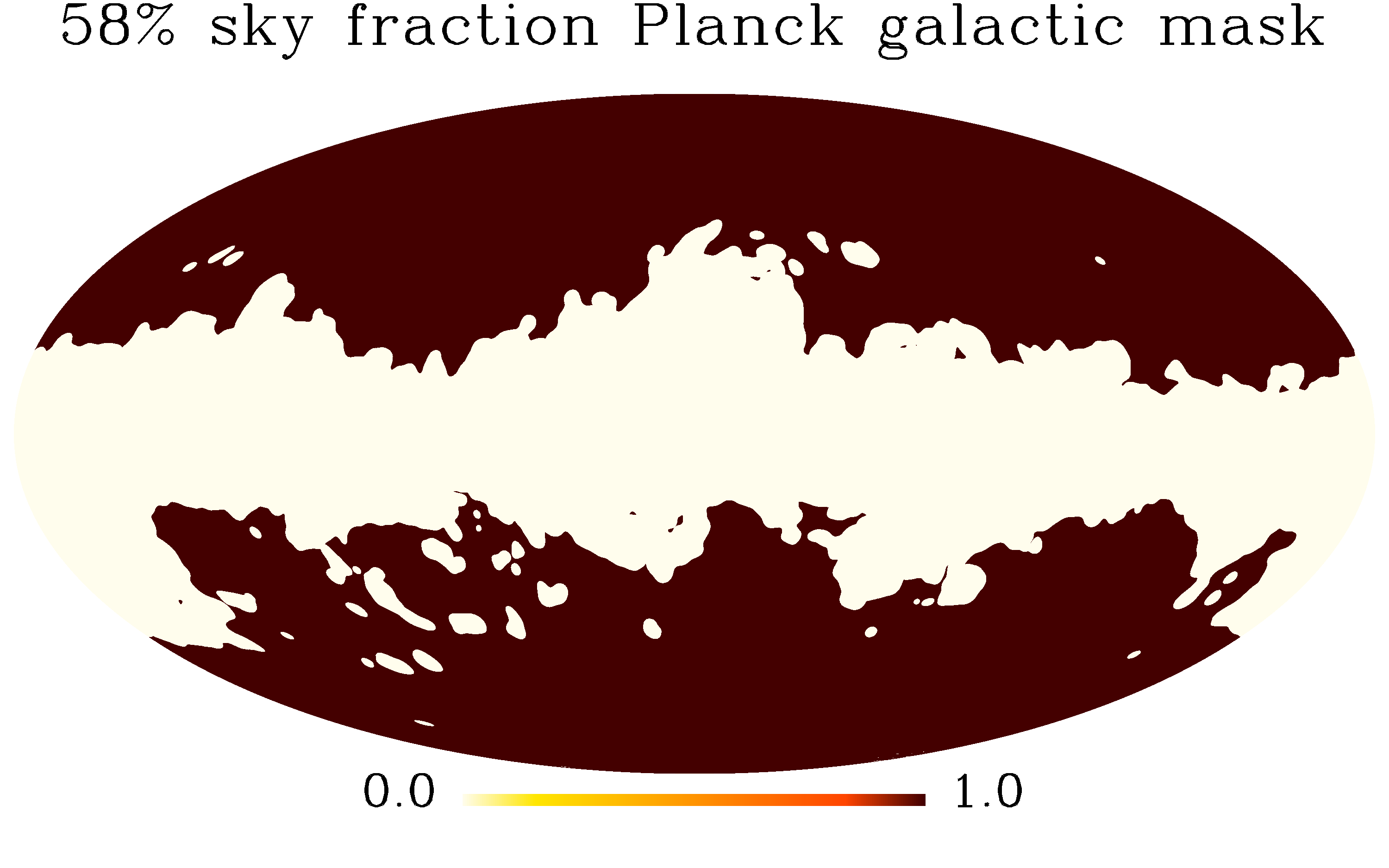}}
\caption{\label{fig:maskplck} Galactic mask released by the Planck
  collaboration with the $y$-distortion maps with $58\%$ unmasked fraction.}
\end{figure}

\begin{figure}
\resizebox{\hsize}{!}{\includegraphics{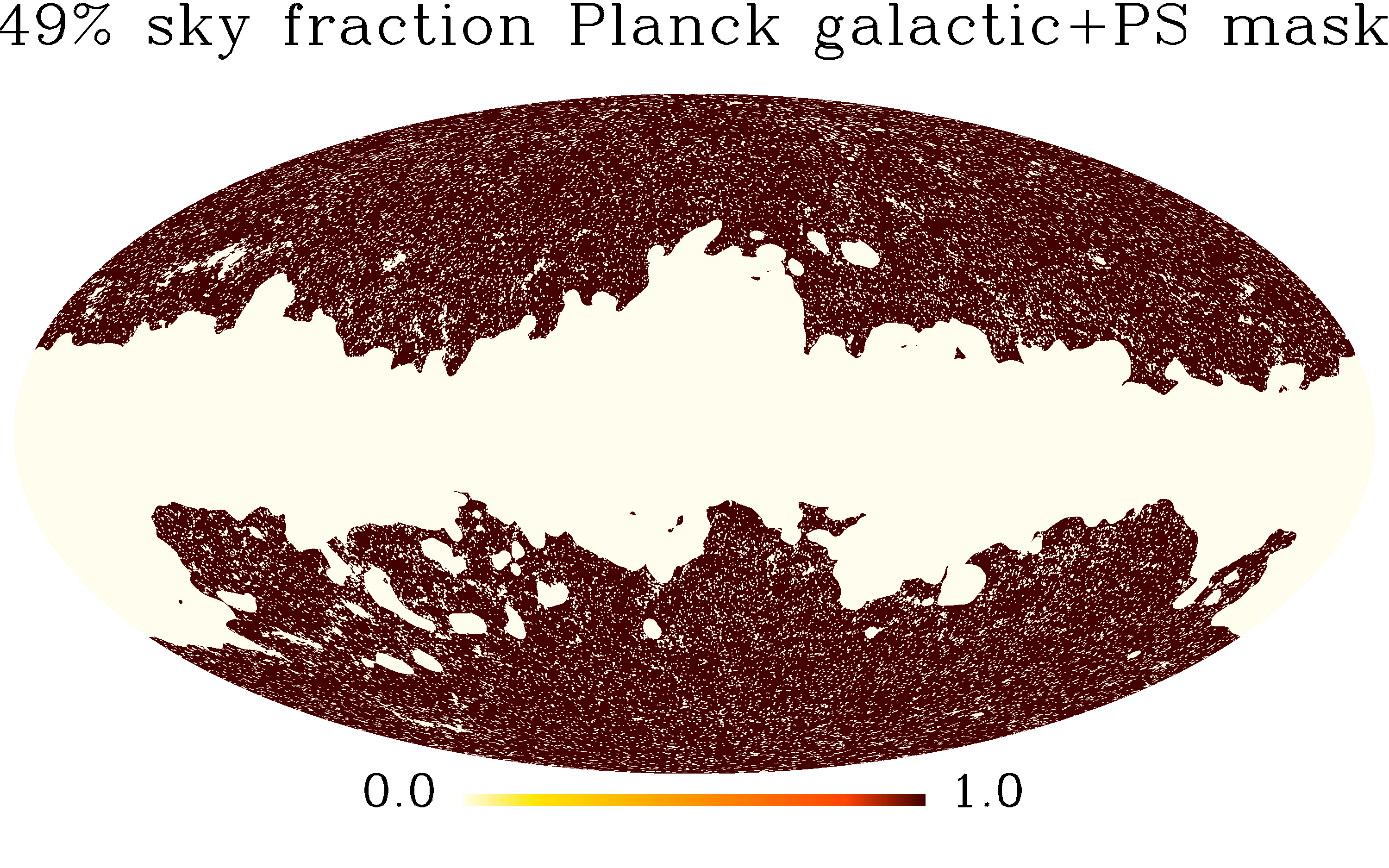}}
\caption{\label{fig:maskplckps}  Planck collaboration point source mask
  added to the galactic mask of Fig. \ref{fig:maskplck}.}
\end{figure}

It is also interesting to compare the performance of our masks with the
galactic and point source masks released by the Planck collaboration with
the $y$-distortion maps.
We show our mask with  $61\%$ unmasked sky fraction, Planck collaboration
galactic mask with $58\%$ sky fraction and with the point source mask added
to this mask yielding $49\%$ sky fraction respectively in
Figs. \ref{fig:masklil}, \ref{fig:maskplck}, \ref{fig:maskplckps}.

\begin{figure}
\resizebox{\hsize}{!}{\includegraphics{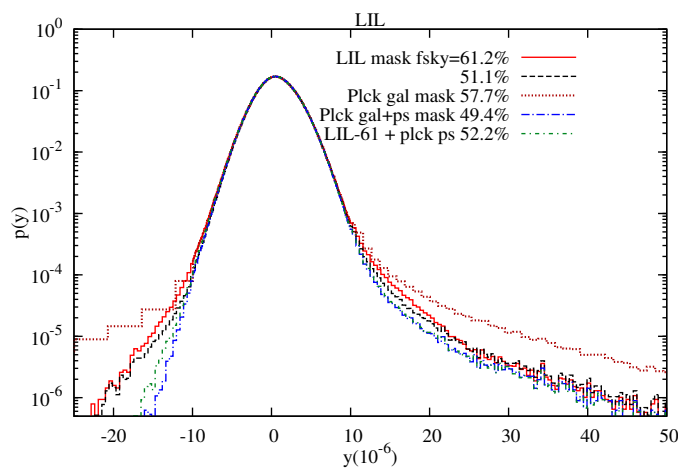}}
\caption{\label{fig:lilpdf} Comparison of different masks for the LIL $y$-map.}
\end{figure}

\begin{figure}
\resizebox{\hsize}{!}{\includegraphics{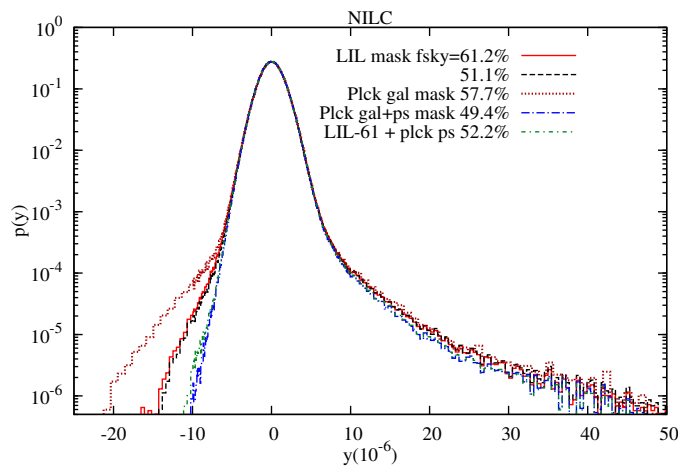}}
\caption{\label{fig:nilcpdf} Comparison of different masks for the NILC $y$-map.}
\end{figure}

Our mask is not much more complicated than the Planck galactic mask but
selectively masks out the point sources and molecular clouds which
contaminate the $y$-distortion map. We plot the PDFs for LIL $61\%$ and
$51\%$ masks as well as for the two Planck masks including and excluding
the point source mask (labeled 'plck') for LIL and NILC $y$-maps in
Figs. \ref{fig:lilpdf} and \ref{fig:nilcpdf} respectively. For NILC
PDF, we see that there are some peaks in the positive tail when using the
Planck galactic mask which are not present in the LIL mask and are because
of the molecular cloud contamination. The contamination in the negative
tail is also significantly less for the LIL mask. The negative tail as well
as the contamination in the positive tail is further reduced when we add
the Planck point source mask. The difference between the Planck galactic
mask and LIL mask is much greater for the LIL $y$-map because of the
molecular cloud regions not covered by the galactic mask. The point source
mask is however very dense and complicated and may also be masking
significant genuine $y$-distortion signal \cite{planckymap}. Our simple mask based on the model
selection approach provides a good compromise between the two extremes of
only using the Planck galactic masks  and Planck galactic +point source
masks. The LIL mask significantly reduces the contamination compared to the
Planck galactic mask but without the overly
complicated structure and complements the masks provided by the Planck collaboration.

\begin{figure}
\resizebox{\hsize}{!}{\includegraphics{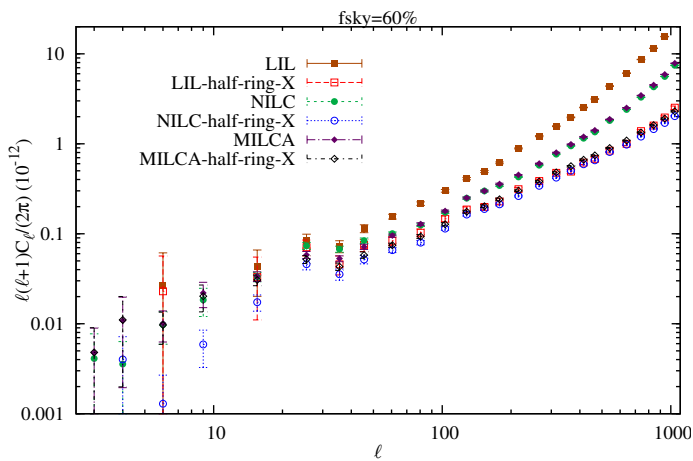}}
\caption{\label{fig:szcl} Comparison of angular power spectra of LIL, MILCA,
  and NILC $y$-maps for $60\%$ sky fraction mask.}
\end{figure}

Finally we compare the angular power spectra of NILC, MILCA, and LIL maps
in Fig. \ref{fig:szcl}. We show the power spectra for the full mission
maps, which includes noise, and the cross-spectra of the half-ring maps in
which the noise is canceled. All power spectra are calculated with
the publicly available PolSpice code \cite{ps1,ps2}  and the effect of the
mask \cite{hivon2002} as well as
the $10'$ beam has been deconvolved. PolSpice also calculates the covariance
matrix \cite{e2004,cc2005,xspect} which we have used to calculate the error
bars for the $\ell$ bins.
For the full mission map, we see the difference between the NILC/MILCA and
LIL maps is because of the higher noise. All power spectra agree when the
noise power is taken out in the half-ring cross spectra (labeled with
half-ring-X suffix). Note that since the ILC algorithms are applied in
small patches on the sky, the large scale modes, $\ell \lesssim 10$, would be influenced by the
details of implementation, e.g. if monopole is subtracted in each patch.
A LIL mask with $60\%$ sky fraction (which is a smoother version of our
$61\%$ mask created using the 545 GHz channel map smoothed with
$15^{\circ}$ FWHM beam combined with the minimal $86\%$ CO mask)  and apodized with a Gaussian in pixel
space was used. We apodized by replacing the $1s$ in the mask by
$1-\exp\left[-9\theta^2/(2\theta_{\rm ap}^2)\right]$, with $\theta_{\rm
  ap}=60'$ except for small point source like regions in the mask where we
used $\theta_{\rm
  ap}=30'$ and $\theta$ is the distance from the mask edge.

\section{Implications for the Planck cluster catalogue}
 Planck collaboration has  released the second cluster
catalogue \citep{planckclusters2015} and we can do additional validation of our algorithm with
this catalogue. {Throughout this paper we  use the second Planck
catalogue.} In particular we can check that the confirmed clusters fall
outside our mask. We can also validate the Planck cluster catalogue using our
results. In particular any unconfirmed candidates in the catalogue which
fall inside our mask are likely to be molecular clouds rather than
clusters.

We were however very conservative in creating our mask and we expect
several weak molecular clouds to remain outside the mask. Instead of
checking 
whether particular  clusters fall inside or outside our mask we follow a
different approach as follows. Since we have positions of cluster
candidates in the catalogue, we can go to the position of each cluster and
use our $\chi^2_{\rm CO-y}$ map to classify it as a cluster or a molecular
cloud. The algorithm we use is the following:

\begin{enumerate}
\item We sum the $\chi^2$ weighted by the square of signal in each pixel
  for all pixels within $5'$ radius of each source in the catalogue and
  divide by the square of the maximum signal. We do
  this for both the CO maps and the $y$-maps to get the biased sum of squares
  $\sum\chi^2_{\rm CO}$ and  $\sum\chi^2_{\rm y}$ for
  each source. This way we give more weight to the pixels at the centre of
  the source where we expect highest S/N and less weight to more numerous
  but low S/N pixels on the outskirts.
\item Since we used model selection with $\Delta \chi^2=3.8$ threshold
  while making the CO and $y$ maps, for the weak sources there will be many
  pixels for which the CO or y amplitude is zero. We penalize these pixels
  by increasing the $\sum \chi^2$, $\sum\chi^2\longrightarrow
  \sum\chi^2+4.0$ for every zero pixel which is surrounded by
  non-zero pixels. This leaves
  out pixels which are zero and also surrounded by zero pixels. We penalize
  the map with smaller number of zero pixels by adding additional penalty
  to the corresponding $\chi^2$ of $4\Delta N$, where $\Delta N$ is the
  difference in zero pixels in the two maps which are also surrounded by
  zero pixels.
\item If the final difference $\Delta (\sum\chi^2)_{\rm CO-y}\equiv\sum\chi^2_{\rm
    CO}-\sum\chi^2_{\rm y} \ge 10.0$, we classify the source as cluster and
  add the annotation $CLG$ to the table, if $\Delta (\sum\chi^2)_{\rm CO-y} \le -10.0$ we
  classify it as a molecular cloud with annotation $MOC$. 
\item For a smaller threshold difference in $\Delta (\sum\chi^2)_{\rm CO-y}$ of $5$, we
  classify the cluster as $pCLG$ or $pMOC$, with $p$ signifying the lower
  significance of classification. Also if one of the maps has number of
  non-zero pixels less than 15, we use the annotation  $pCLG$ or $pMOC$
  even if the  $|\Delta (\sum\chi^2)_{\rm CO-y} | > 10.0$. 
\item In the later case
  for the  $\sum \chi^2  $ of the map with higher number of
  non-zero pixels $N$, if $1.6 N < \sum \chi^2  $ we classify the
  source as $IND$. If the difference $\Delta (\sum\chi^2)_{\rm CO-y}$ is below the threshold we
  classify the source as $IND$ or indeterminable by our algorithm. If both
  maps have non-zero pixels less than 15 we classify the source as $IND$.
\end{enumerate}

This mostly heuristic algorithm works quite well in practice and  we
verified it
by looking at the confirmed sources. There is in general a big difference
in $\chi^2$ for strong sources unambiguously determining whether a source is
a cluster or a molecular cloud. The classification we get is not very
sensitive to small changes in the thresholds. Our full annotated Planck
cluster catalogue is available at \url{http://theory.tifr.res.in/~khatri/szresults/}. Note that we retain all entries
from the Planck official catalogue and just add additional column with our
annotation. We reproduce the top 10 highest signal to noise clusters from
the catalogue in Table \ref{Tbl:cluster} and top highest signal to noise
sources as well as 10 lowest signal to noise sources but with $S/N>6$ that we have classified as molecular clouds in \ref{Tbl:moc}.

\begin{table*}
\caption{\label{Tbl:cluster}Highest S/N noise clusters in the Planck
  catalogue. All columns except for $\Delta (\sum\chi^2)_{\rm
     CO-y}$ and validation columns are from the Planck catalogue. The
  key to Planck validation column can be found in \citep{planckclusters2015} and tells which
  external source was use for validation, a $-1$ in this column mean an
  unconfirmed candidate. The last columns tells whether this source is
  included in the sample used for cosmological analysis. Also shown is the
  quality factor from Planck catalogue, $Q_N$ \citep{agha2014} with $Q_N>0.4$
  suggested by the Planck collaboration as the criterion to describe the
  good quality clusters. Full version of Tables \ref{Tbl:cluster},
  \ref{Tbl:moc}  and \ref{Tbl:Qn}  is  available in electronic form
at the CDS via anonymous ftp to \url{cdsarc.u-strasbg.fr} (130.79.128.5)
or via \url{http://cdsweb.u-strasbg.fr/cgi-bin/qcat?J/A+A/}}
\begin{tabular}{|c|c|c|c|c|c|c|c|}
\hline
   cluster & S/N&z&Planck valid.&$\Delta (\sum\chi^2)_{\rm
     CO-y}$ & validation &cosmology & $Q_N$\\
\hline
PSZ2 G075.71+13.51 & 48.98511 &  0.05570  &  21   &   893.456 & CLG & T & 0.994\\
PSZ2 G110.98+31.73 & 40.75489 &  0.05810  &  21   &   294.893 & CLG & T & 0.992\\
PSZ2 G272.08-40.16 & 39.99466 &  0.05890  &  21   &   492.870 & CLG & T & 0.993\\
PSZ2 G239.29+24.75 & 36.24374 &  0.05420  &  21   &   192.400 & CLG & T & 0.993\\
PSZ2 G057.80+88.00 & 35.69822 &  0.02310  &  21   &   418.131 & CLG & T & 0.992\\
PSZ2 G006.76+30.45 & 35.01054 &  0.20300  &  21   &   137.806 & CLG & F & 0.994\\
PSZ2 G324.59-11.52 & 32.40285 &  0.05080  &  21   &   321.450 & CLG & F & 0.993\\
PSZ2 G044.20+48.66 & 28.38608 &  0.08940  &  21   &   127.431 & CLG & T & 0.994\\
PSZ2 G266.04-21.25 & 28.38260 &  0.29650  &  21   &   103.555 & CLG & T & 0.993\\
PSZ2 G072.62+41.46 & 27.43035 &  0.22800  &  20   &    88.383 & CLG & T & 0.994\\
\hline
\end{tabular}
\end{table*}

\begin{table*}
\caption{\label{Tbl:moc}Sample of sources from the Planck catalogue
  classified as molecular clouds, columns are same as in Table \ref{Tbl:cluster}.} 
\begin{tabular}{|c|c|c|c|c|c|c|c|}
\hline
   cluster & S/N&z&Planck valid.&$\Delta (\sum\chi^2)_{\rm
     CO-y}$ & validation &cosmology& $Q_N$\\
\hline
 PSZ2 G153.56+36.82 &  15.89673 & -1.00000  &   -1 &  -528.090 & MOC & F& 0.000\\
 PSZ2 G182.42-28.28 &  15.77494 &  0.08820  &   21 &   -15.384 & MOC & F& 0.991\\
 PSZ2 G342.45+24.14 &  15.71413 & -1.00000  &   -1 & -2194.689 & MOC & F& 0.035\\
 PSZ2 G284.97-23.69 &  15.65867 &  0.39000  &   20 &   -58.154 & MOC & F& 0.991\\
 PSZ2 G314.96+10.06 &  15.49399 &  0.09660  &   21 &   -35.386 & MOC & F& 0.990\\
 PSZ2 G171.98-40.66 &  13.39432 &  0.27000  &   20 &   -53.838 & MOC & F& 0.964\\
 PSZ2 G125.37-08.67 &  12.29307 &  0.10660  &   21 &   -30.983 & MOC & F& 0.974\\
 PSZ2 G100.45+16.79 &  11.78533 & -1.00000  &   -1 & -7597.947 & MOC & F& 0.024\\
 PSZ2 G105.82-38.36 &  11.51047 & -1.00000  &   15 &  -342.830 & MOC & F& 0.000\\
 PSZ2 G340.09+22.89 &  11.35395 & -1.00000  &   -1 & -2443.363 & MOC & F& 0.033\\
 PSZ2 G338.04+23.65 &   6.05953 & -1.00000  &   -1 & -1315.602 & MOC & F& 0.034\\
 PSZ2 G028.08+10.79 &   6.03667 &  0.08820  &   21 &  -119.810 & MOC & F& 0.875\\
 PSZ2 G093.04-32.38 &   6.03185 & -1.00000  &   -1 &  -370.231 & MOC & F& 0.006\\
 PSZ2 G337.95+22.70 &   6.03163 & -1.00000  &   -1 & -1959.108 & MOC & F& 0.047\\
 PSZ2 G278.74-45.26 &   6.03076 & -1.00000  &   16 &   -67.508 &pMOC & F& 0.002\\
 PSZ2 G198.73+13.34 &   6.02919 & -1.00000  &   20 &   -51.949 & MOC & F& 0.311\\
 PSZ2 G215.24-26.10 &   6.02551 &  0.33600  &   21 &   -10.723 & MOC & F& 0.993\\
 PSZ2 G299.54+17.83 &   6.02125 & -1.00000  &   -1 &   -27.199 & MOC & T& 0.983\\
 PSZ2 G076.44+23.53 &   6.01971 &  0.16900  &   11 &    -6.638 &pMOC & T& 0.967\\
 PSZ2 G281.26-46.90 &   6.00791 &  0.28400  &   20 &    -5.943 &pMOC & T& 0.998\\
\hline
\end{tabular}
\end{table*}

\begin{figure*}
\resizebox{12 cm}{!}{\includegraphics{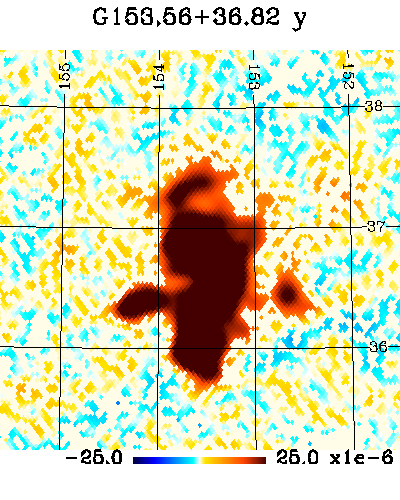}\hspace{10
  pt}\includegraphics{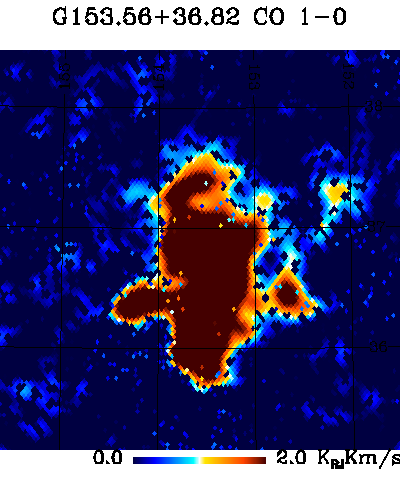}\hspace{10
  pt}\includegraphics{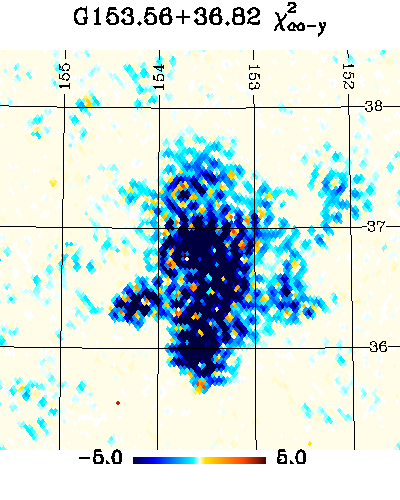}}\\
\resizebox{12 cm}{!}{\includegraphics{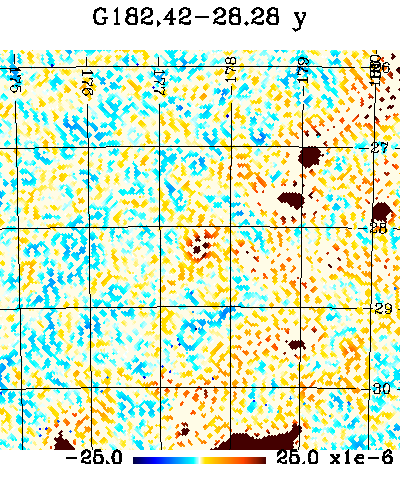}\hspace{10
  pt}\includegraphics{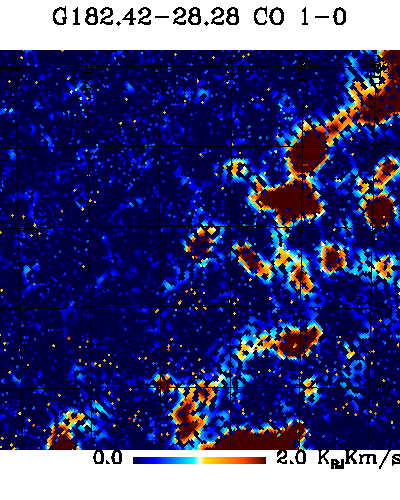}\hspace{10
  pt}\includegraphics{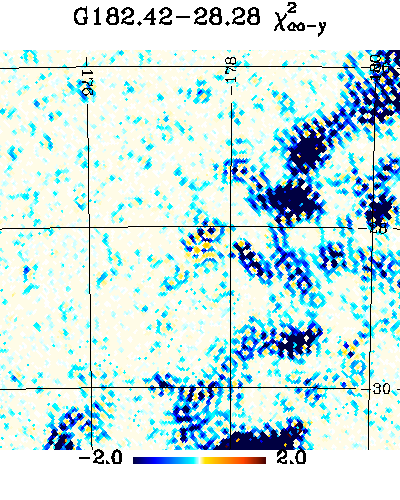}}\\
\resizebox{12 cm}{!}{\includegraphics{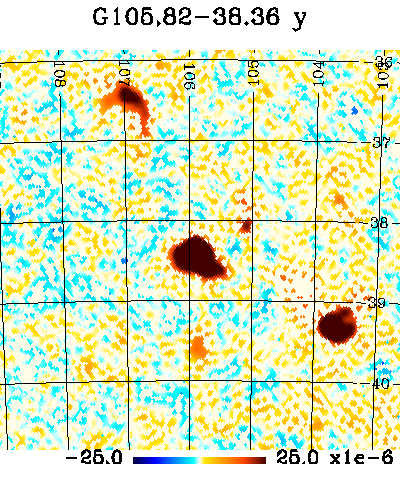}\hspace{10
  pt}\includegraphics{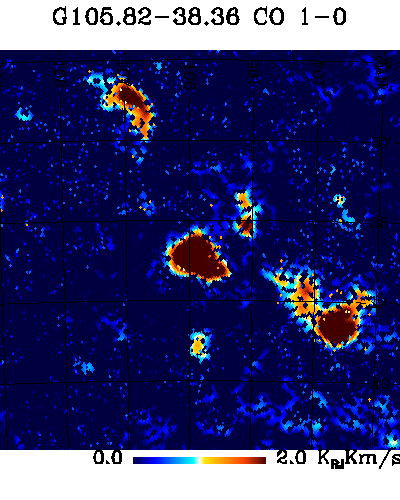}\hspace{10
  pt}\includegraphics{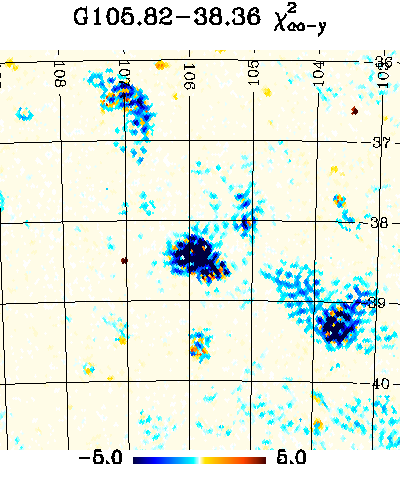}}\\
\sidecaption
\resizebox{12 cm}{!}{\includegraphics{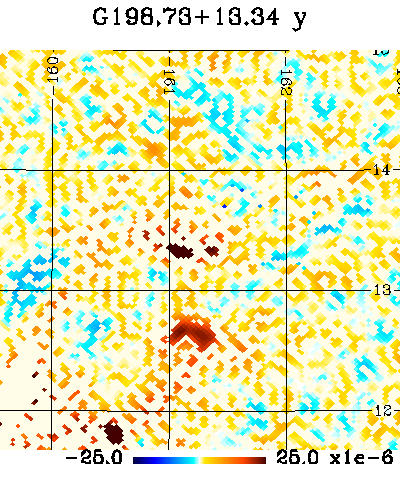}\hspace{10
  pt}\includegraphics{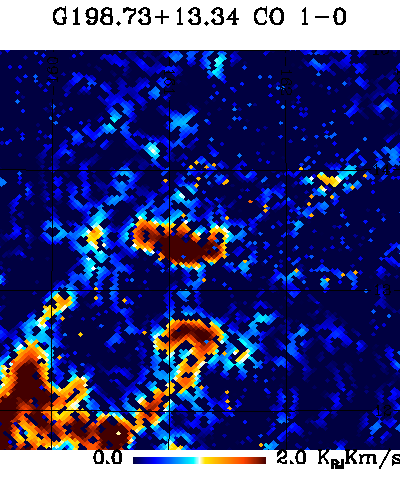}\hspace{10
  pt}\includegraphics{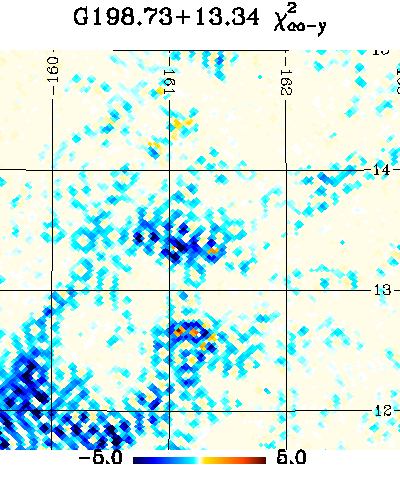}}
\caption{\label{Fig:moc} Some of the candidates from Table \ref{Tbl:moc}
  identified by our algorithm as molecular clouds or having significant
  contamination from the molecular clouds.}
\end{figure*}

\begin{figure*}
\resizebox{\hsize}{!}{\includegraphics{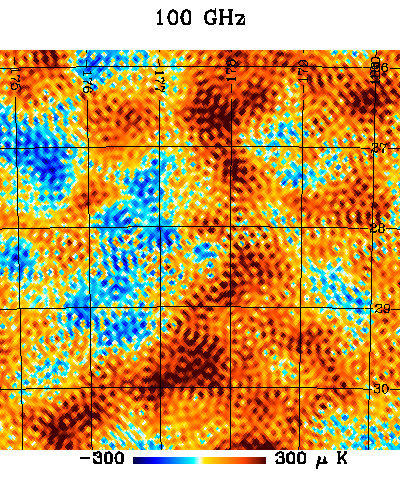}\hspace{10
  pt}\includegraphics{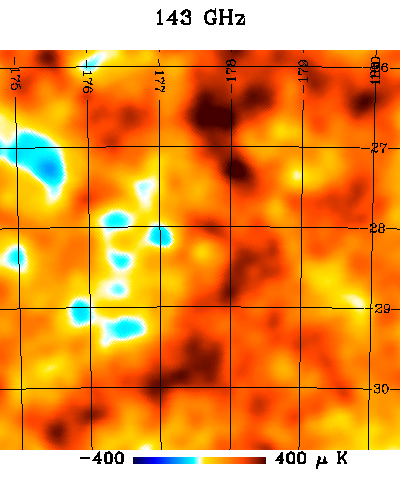}\hspace{10
  pt}\includegraphics{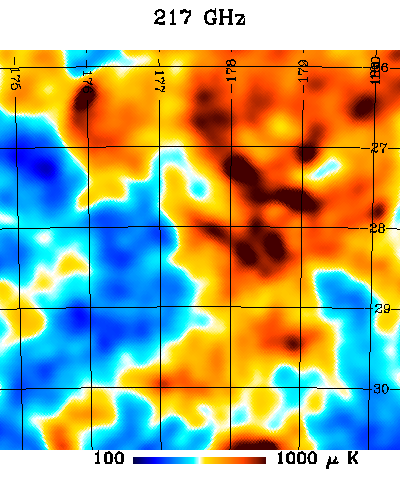}\hspace{10
  pt}\includegraphics{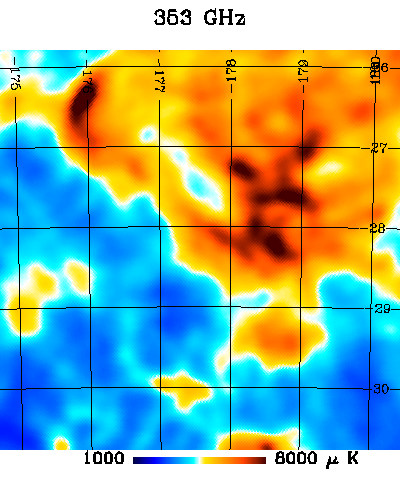}}
\caption{\label{Fig:moc2} Original Planck full channel maps smoothed to 10'
resolution at the location of the second source in Fig. \ref{Fig:moc}
confirm the presence of molecular clouds in the foreground. Note that the
scales in the last two maps are strictly positive.}
\end{figure*}

It may seem surprising that quite a few confirmed clusters are identified
as molecular clouds and this needs some explanation. First we note that
the unconfirmed clusters have extremely high $\Delta (\sum \chi^2)_{\rm CO-y}$
values and it would be very surprising if they turned out to be actual clusters. We show in Fig. \ref{Fig:moc}
some sources from Table \ref{Tbl:moc} with different Planck validation
flags. 
The first source is an unconfirmed candidate that is almost surely a
molecular cloud according to our analysis. This is also obvious  visually 
in Fig. \ref{Fig:moc} The second row in Fig. \ref{Fig:moc} requires more
attention. The source is located at the centre of the frame and is at the
edge of a large molecular cloud. There is a confirmed x-ray cluster in the
MCXC catalogue (meta-xatalogue of X-ray detected clusters of galaxies \citep{mcxc}) at this
position and our $\chi^2$ map shows a few pixels where a cluster is
preferred over molecular cloud. However there is still significant CO
emission from the foreground molecular cloud which seems to dominate over
the $y$-signal and we expect that the amplitude of the distortion signal in the
$y$-map to be unreliable. This interpretation is also confirmed by looking at the
individual frequency maps centred at the same location in Fig. \ref{Fig:moc2}.

The third source in Fig. \ref{Fig:moc} was confirmed by arc-minute
micro-Kelvin interferometer (AMI) \citep{ami} which operates at very low
frequency of 15 GHz and may have contamination from the synchrotron
emission from the molecular clouds. Our $\chi^2$ test shows that there is a
molecular cloud clump at this location. The last row in Fig. \ref{Fig:moc}
shows a source with low quoted S/N of $\approx 6$ in the Planck catalogue
which has Planck validation flag of 20 implying that it was present in the
2013 Planck catalogue \citep{planckclusters} where it is listed as confirmed 
but there is no redshift information. Even if there is a cluster behind the
molecular cloud, our analysis shows significant CO contamination to the
$y$-signal. 

\begin{figure*}
\resizebox{\hsize}{!}{\includegraphics{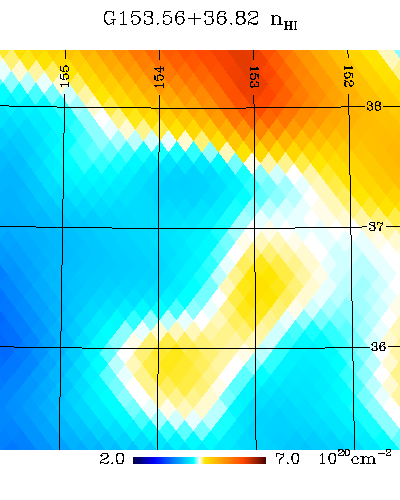}\hspace{10
  pt}\includegraphics{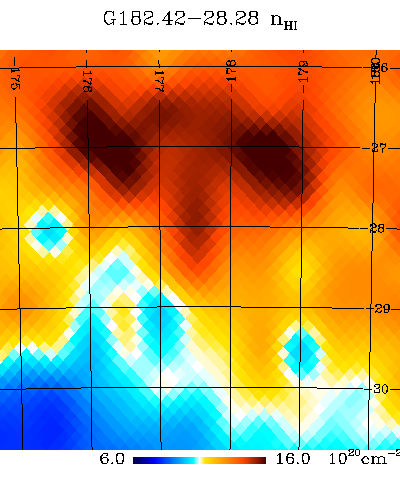}\hspace{10
  pt}\includegraphics{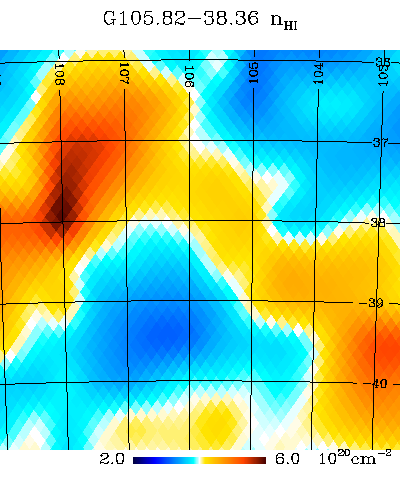}\hspace{10
  pt}\includegraphics{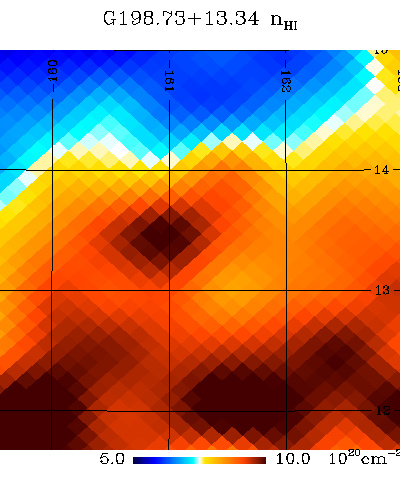}}\\
\resizebox{\hsize}{!}{\includegraphics{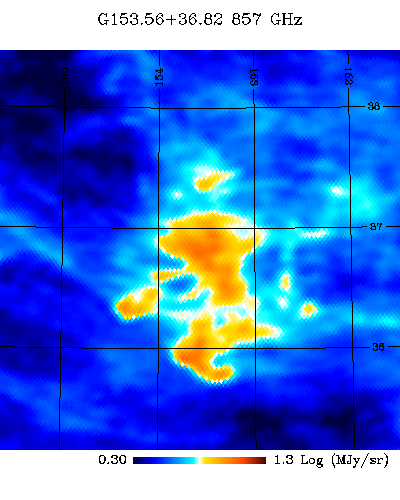}\hspace{10
  pt}\includegraphics{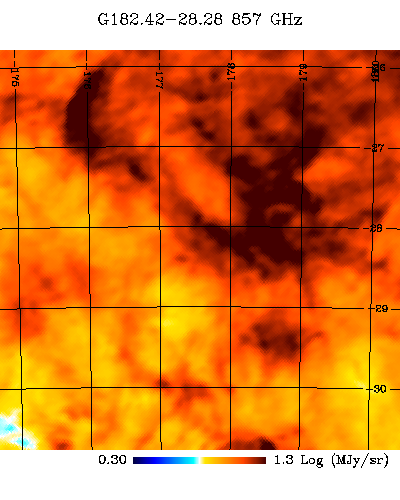}\hspace{10
  pt}\includegraphics{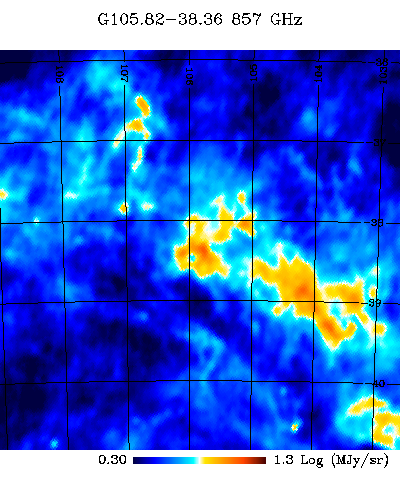}\hspace{10
  pt}\includegraphics{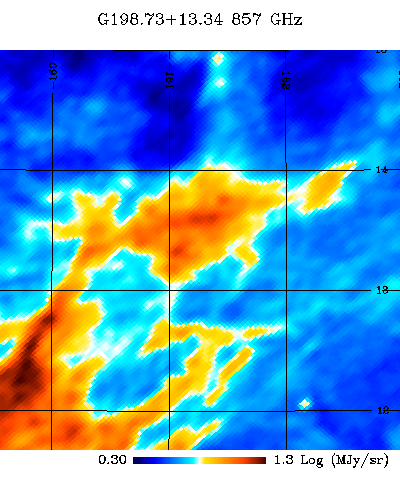}}
\caption{\label{Fig:HIdust} HI column density and Planck 857 GHz emission
  for the sources shown in Fig. \ref{Fig:moc}. The scale of all figures is same
as the respective figures in Fig. \ref{Fig:moc} with the cluster candidates
from the Planck catalogue at the centre of the frame.}
\end{figure*}

{In top panel in Fig. \ref{Fig:HIdust} we show the HI column density maps
calculated by combining the Leiden/Dwingeloo survey data \citep{hkb1996,HB1997} and
the composite HI column density map of \citet{dl1990} and publicly
available at
\url{http://lambda.gsfc.nasa.gov/product/foreground/combnh_map.cfm} for the
same sources as Fig. \ref{Fig:moc}. In the bottom panel the Planck 857 data
for the respective sources is shown. For the first source, the location of
the cluster candidate coincides with region of low HI column density
surrounded by high HI column density. The  dust emission in the 857 GHz
channel is also higher in the location of the source than in the
surroundings. This is the signature that we would expect from a molecular
cloud surrounded by an envelop of atomic HI gas. For the second source the
situation is similar but not as clearly visible because both the dust
 emission and HI column density are higher and we are probably at the edge
 of a bigger molecular cloud. In the third and fourth source again there is
 agreement with the signature of the presence of a molecular cloud. The external HI
 column density data combined with Planck 857 GHz data therefore gives us
 confidence in our interpretation of these sources as molecular cloud candidates.}

\begin{figure*}
\sidecaption
\resizebox{12 cm}{!}{\includegraphics{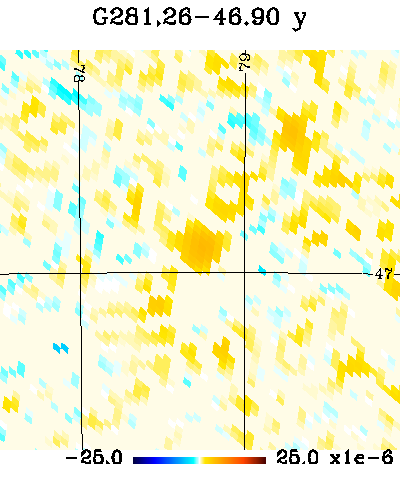}\hspace{10
  pt}\includegraphics{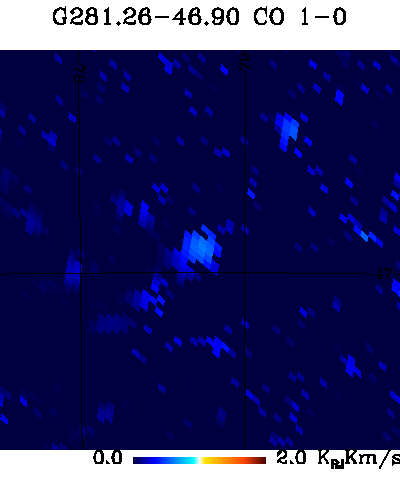}\hspace{10
  pt}\includegraphics{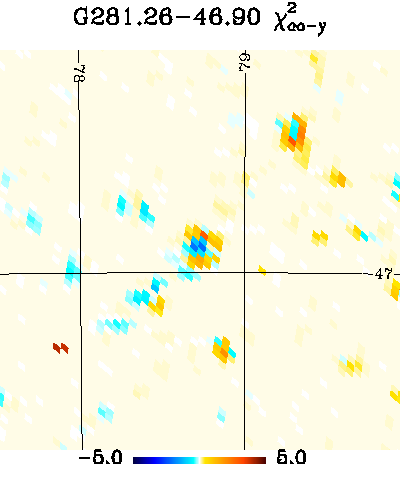}}
\caption{\label{Fig:moc5} One low S/N case where our test becomes unreliable
   Visually however this preference towards it being a cluster can be inferred.}
\end{figure*}

Finally in Fig. \ref{Fig:moc5} we show a marginal case of a low
S/N source, the last entry in Table \ref{Tbl:moc} where there is a confirmed
x-ray cluster but the S/N is too low for a reliable distinction using our test.
This is mostly because of the small angular size of the source resulting in most
  pixels in the $5'$ radius having a zero signal so that our algorithm for
  summing up the $\chi^2$ is not optimal. However visual inspection shows
  that most pixels favor $y$-distortion over the CO contamination. Most of
  such sources would show up in our annotation with weaker validation code $pMOC$
  or $pCLG$.

To summarize, out of a total of 1653 sources in the Planck catalogue, we identify 395 molecular
cloud candidates with annotation 'MOC' and 97 molecular candidates with lower
significance  annotation 'pMOC'. There are 450 unconfirmed clusters
with Planck validation flag '-1', of which we classify 232 or $52\%$
cluster candidates as
'MOC' , and 35 candidates as 'pMOC'. Therefore almost $59\%$ of all the
sources we identify as 'MOC' are unconfirmed candidates. 

\begin{figure*}
\sidecaption
\resizebox{\hsize}{!}{\includegraphics{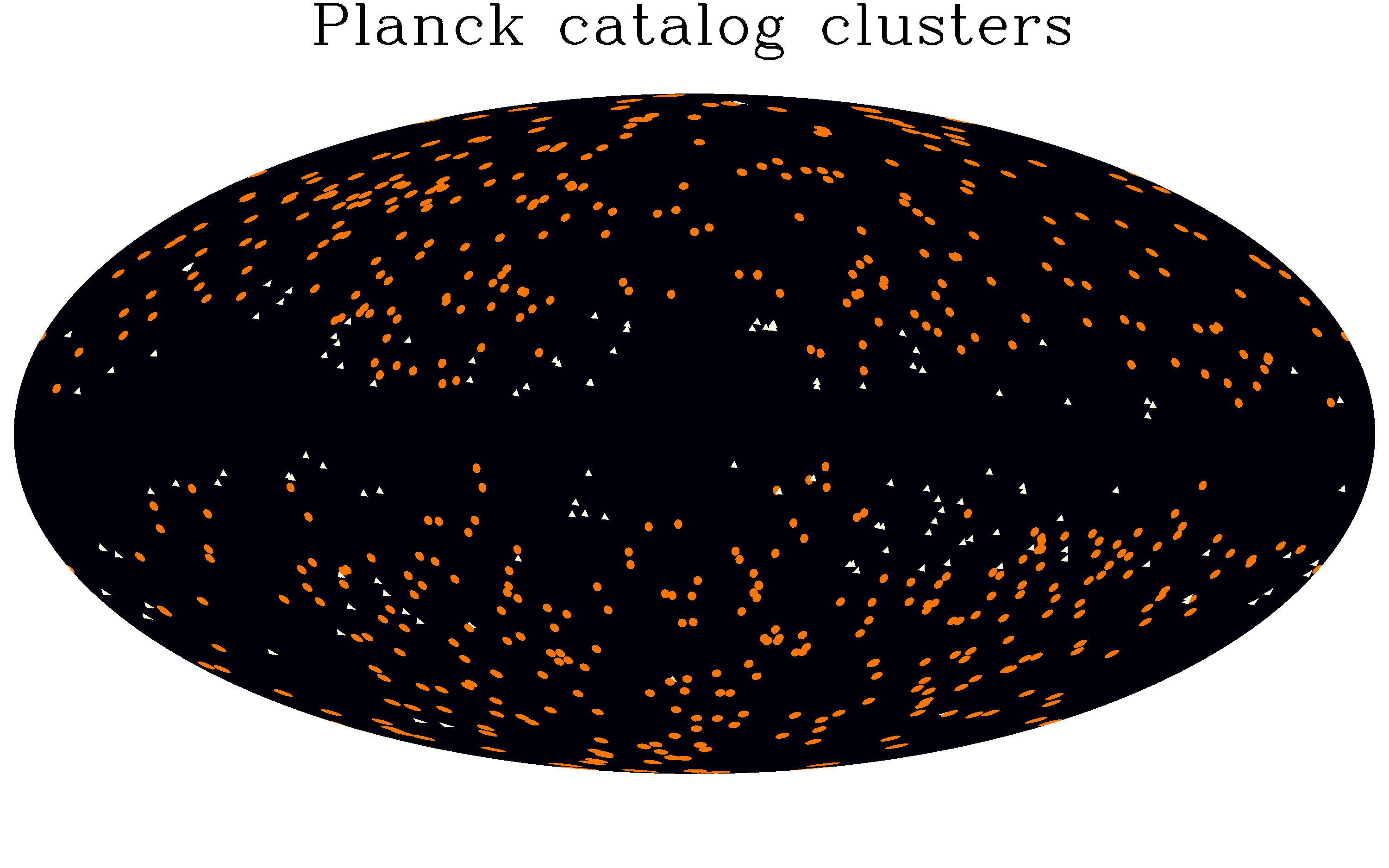}}
\caption{\label{Fig:clgmoc} Position on the sky in galactic coordinates of 528 sources from the Planck catalogue
 identified by us as likely to be clusters or indeterminable (orange circles) and
 130 sources strongly likely
to be molecular clouds (white triangles).}
\end{figure*}

The number of clusters above the threshold used by the Planck collaboration
for cosmology sample, ${\rm S/N}>6$, is 693. Out of these 507 clusters have
the cosmology flag set to 'T'. From this cosmology sample we identify only
19 clusters as 'MOC' and 18 clusters as 'pMOC'. Of the remaining 186
clusters with ${\rm S/N}>6$, we identify 111 sources as 'MOC' and 17
sources as 'pMOC'. We thus confirm that the cosmology sample is much
cleaner and reliable compared to the full cluster catalogue and we should
expect significant contamination from the molecular clouds in the rest of
the catalogue. {We show in Fig. \ref{Fig:clgmoc} the position of the
  130 cluster candidates identified by us as strongly as molecular clouds with annotation
  'MOC' (white triangles) together with 528 sources which are identified by us as clusters or are
  undetermined by our algorithm with our
  annotations 'CLG', 'pCLG' or 'IND' (orange circles).}

\subsection{Comparison with neural network based approach}
\begin{figure}
\resizebox{\hsize}{!}{\includegraphics{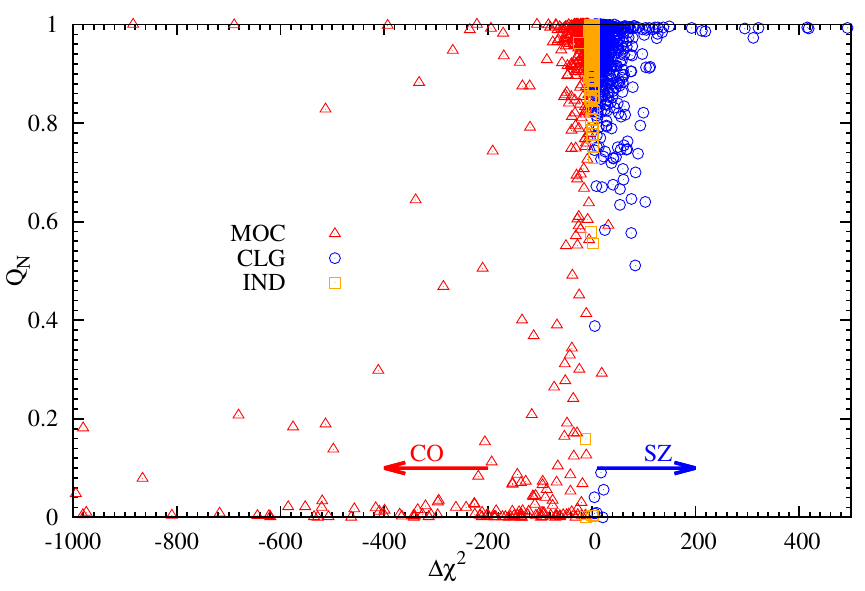}}
\caption{\label{Fig:Qn}Comparison of the two quality measures: our model
  selection based $\Delta (\sum
  \chi^2)_{\rm CO-y}$ and $Q_N$ from \cite{planckclusters2015} based on neural network based approach.}
\end{figure}
\begin{figure*}
\resizebox{\hsize}{!}{\includegraphics{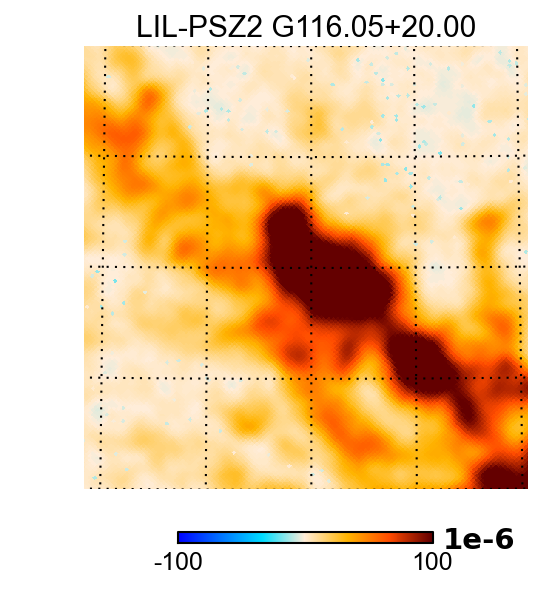}\hspace{10
  pt}\includegraphics{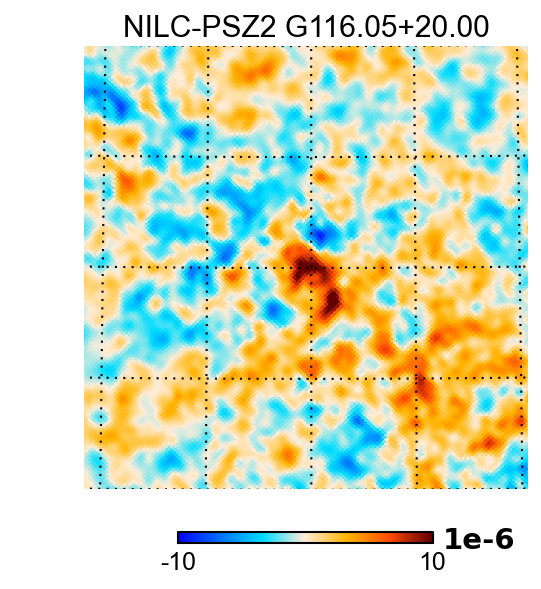}\hspace{10
  pt}\includegraphics{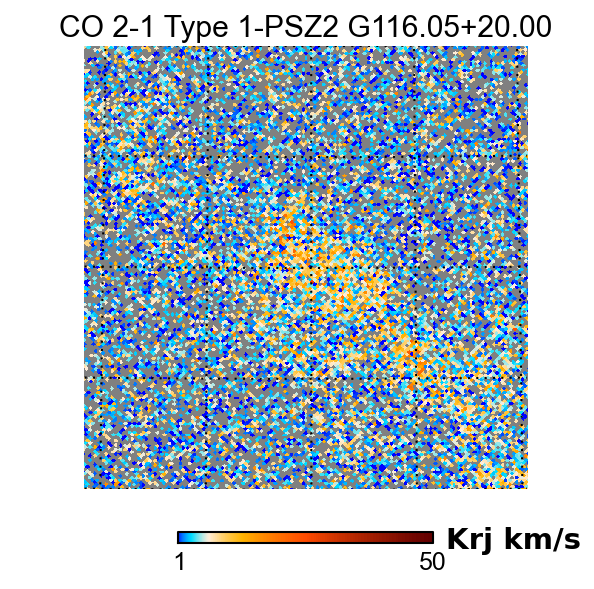}\hspace{10
  pt}\includegraphics{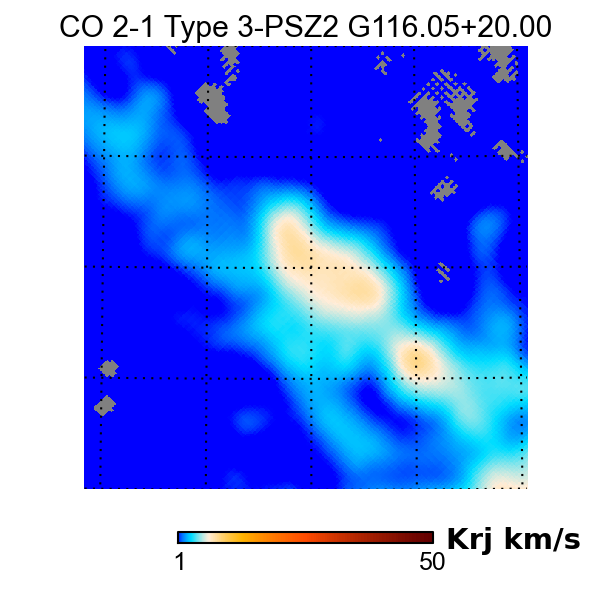}}\\
\resizebox{\hsize}{!}{\includegraphics{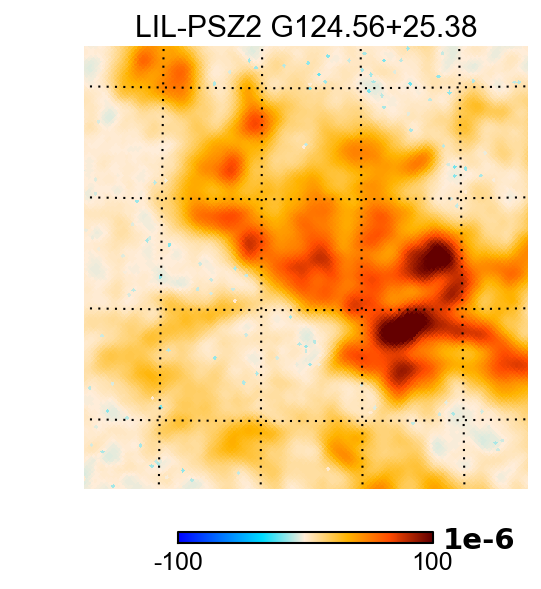}\hspace{10
  pt}\includegraphics{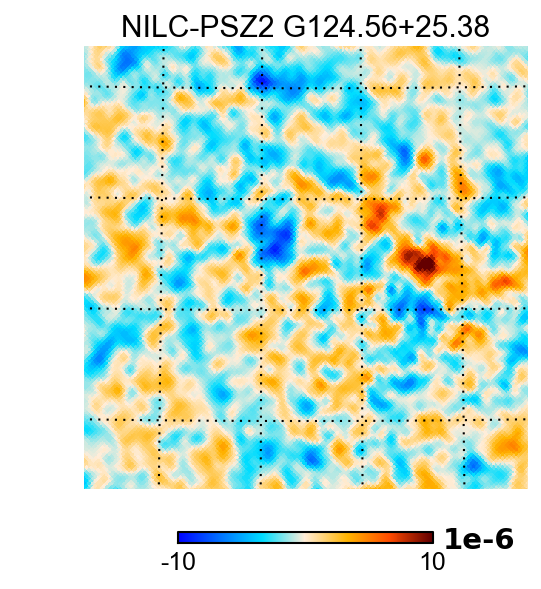}\hspace{10
  pt}\includegraphics{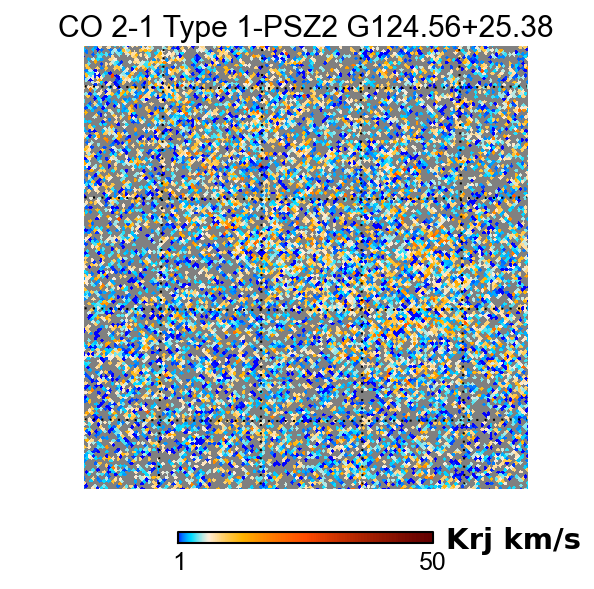}\hspace{10
  pt}\includegraphics{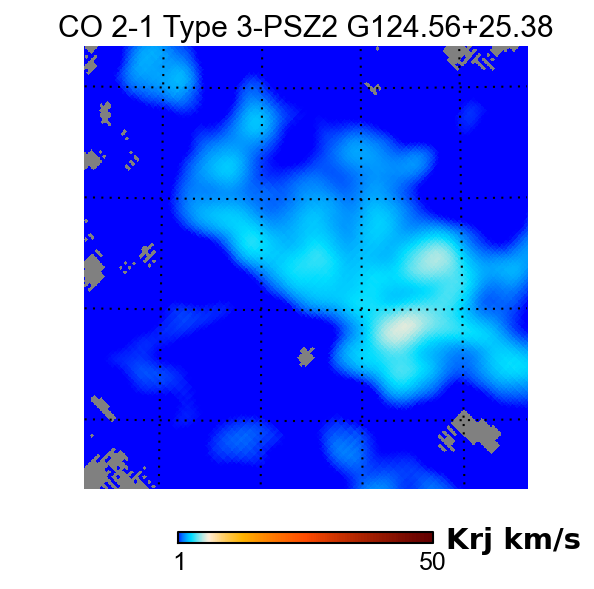}}\\
\resizebox{\hsize}{!}{\includegraphics{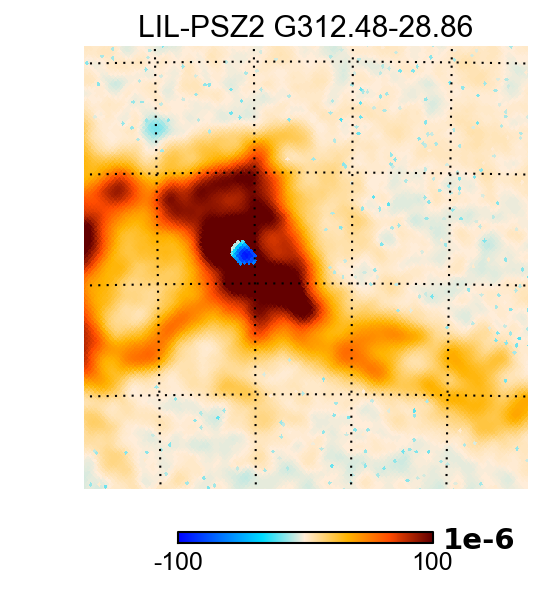}\hspace{10
  pt}\includegraphics{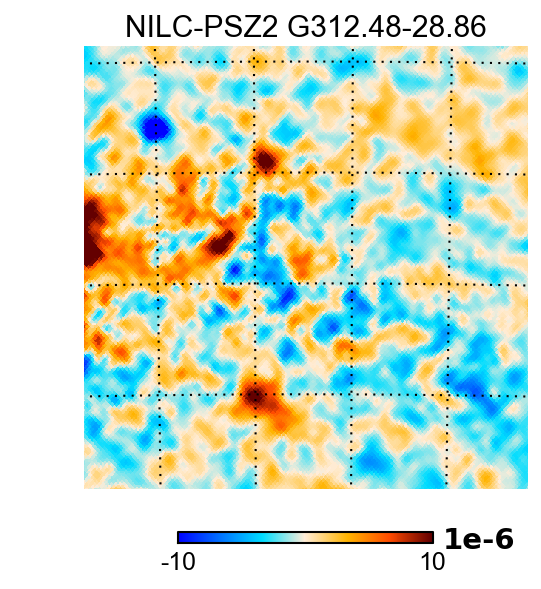}\hspace{10
  pt}\includegraphics{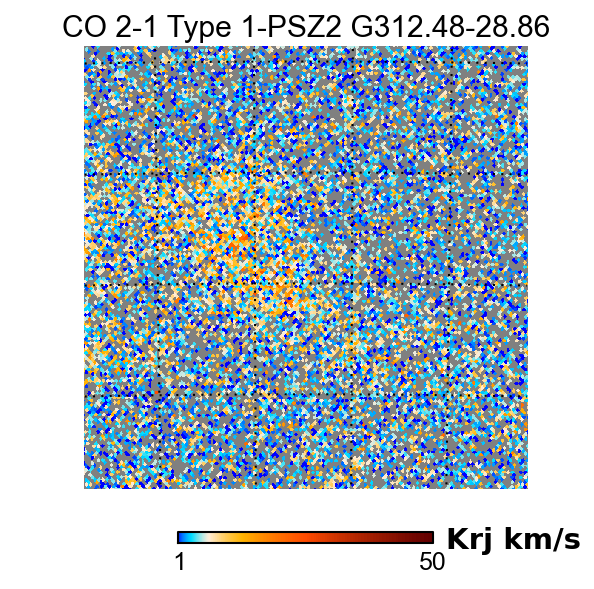}\hspace{10
  pt}\includegraphics{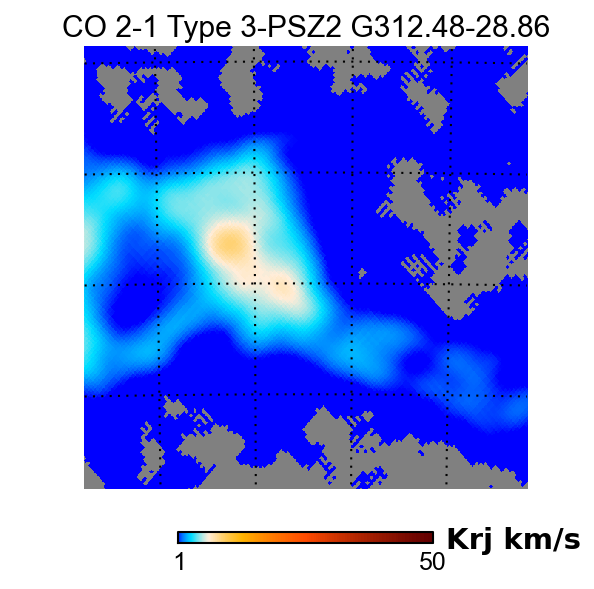}}\\
\resizebox{\hsize}{!}{\includegraphics{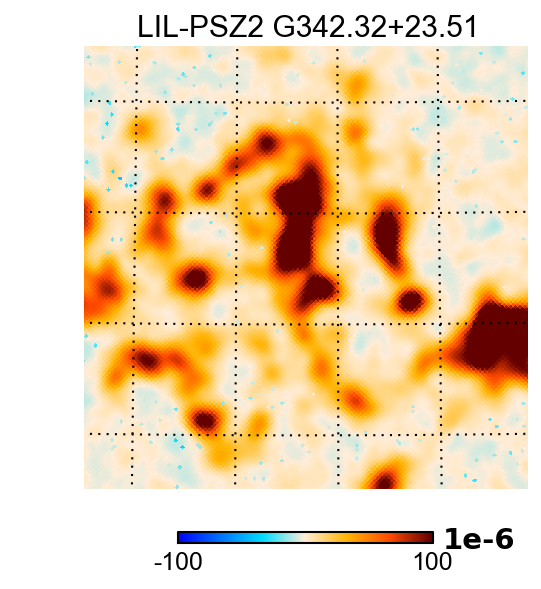}\hspace{10
  pt}\includegraphics{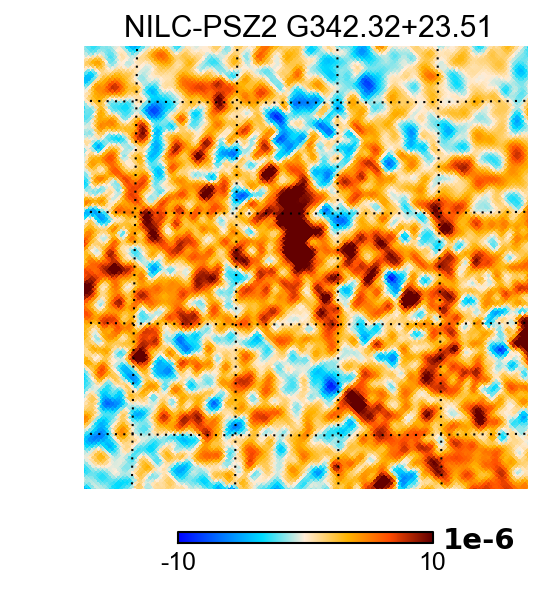}\hspace{10
  pt}\includegraphics{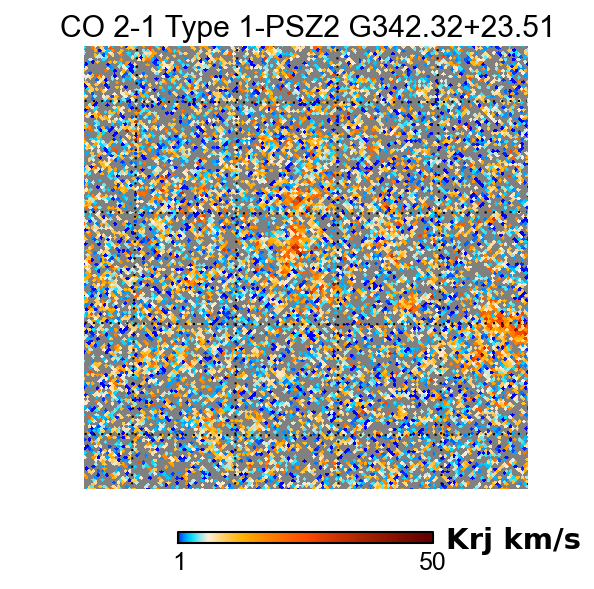}\hspace{10
  pt}\includegraphics{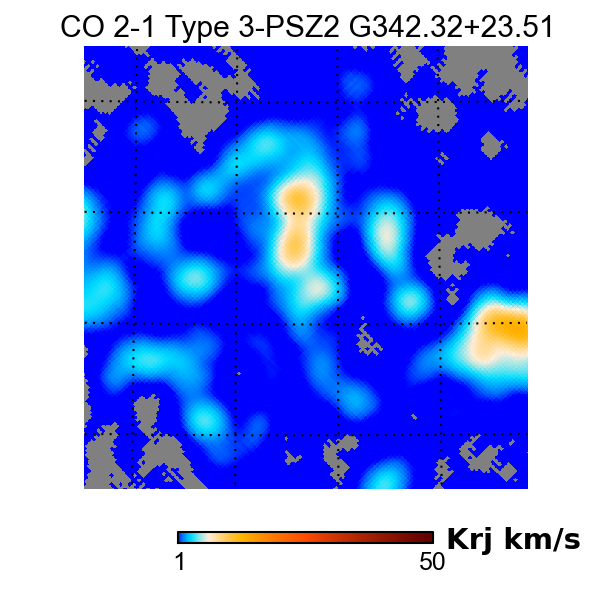}}\\
\resizebox{\hsize}{!}{\includegraphics{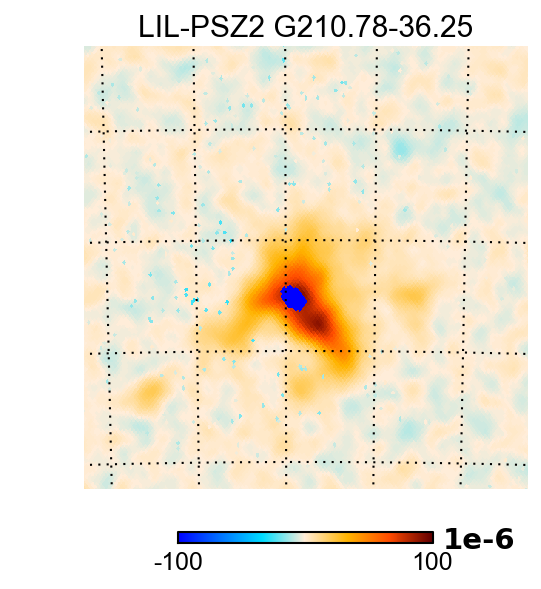}\hspace{10
  pt}\includegraphics{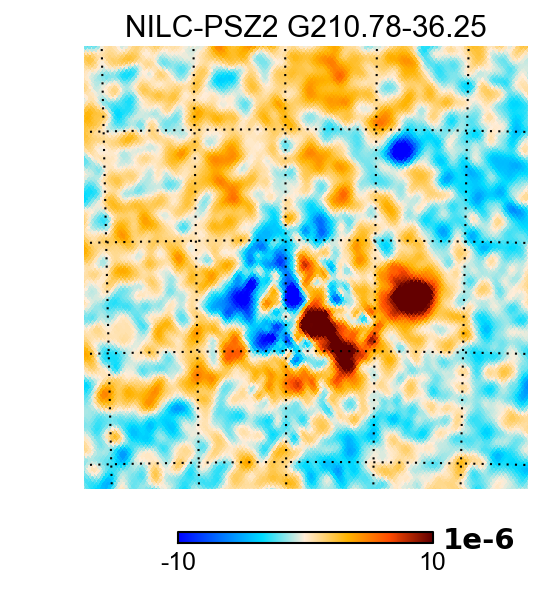}\hspace{10
  pt}\includegraphics{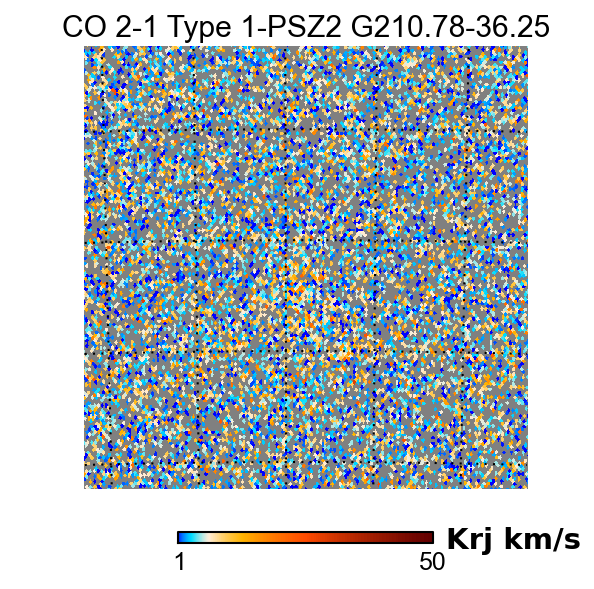}\hspace{10
  pt}\includegraphics{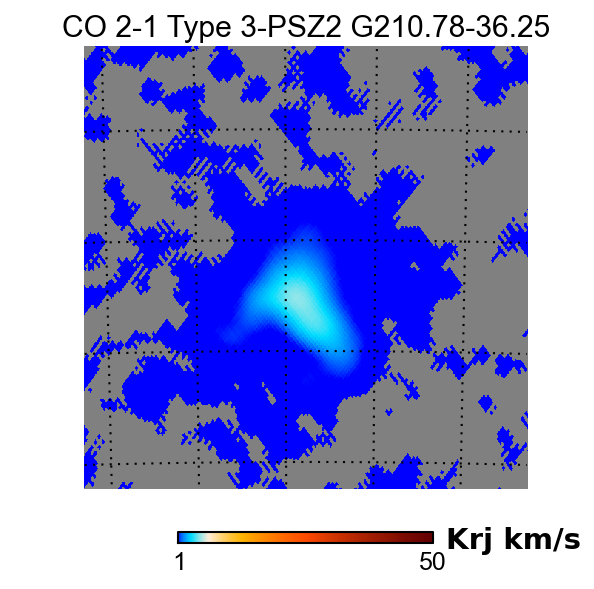}}\\
\caption{\label{Fig:Qnmap} We show the $4^{\circ}\times 4^{\circ}$ regions
  centred around the sources listed in Table \ref{Tbl:Qn} in LIL and NILC $y$ and
  Type 1 and Type 2 CO maps from 2015 release \citep{planckco}. }
\end{figure*}

{We show in Fig. \ref{Fig:Qn} the plot of our $\Delta (\sum
  \chi^2)_{\rm CO-y}$ vs $Q_N$ from \citet{agha2014,planckclusters2015}. Overall
  there is weak correlation between the two measures of quality.
The clusters identified by \cite{planckclusters2015} as very low quality are also strongly
identified by our algorithm as molecular clouds with very large negative values of
$\Delta (\sum \chi^2)_{\rm CO-y}$ in the bottom right quadrant of the plot. Thus for the worst candidates we agree
with \citet{agha2014} and \citet{planckclusters2015}. There is however
strong disagreement for some candidates with out algorithm identifying them
as strongly contaminated while they are given a $Q_N$ close to $1$ in the
Planck catalogue. We list some of the most extreme cases in Table
\ref{Tbl:Qn}. We show LIL and NILC $y$ maps as well as Planck Type 1 and
Type 2 CO maps for these regions in Fig. \ref{Fig:Qnmap}. The Type 1 maps
are noisier but uncontaminated by the SZ signal while the Type maps have
less noise but may be contaminated by the SZ signal leakage \citep{planckco}.  In the first
second and fourth source the location of the cluster coincides with a local
peak in the CO emission maps as  can be seen in the LIL map as well as the
Planck CO emission maps. The third and fifth sources have the source
location at the edge of the molecular cloud. In the NILC map, on which the
cluster catalogue is based, the surrounding molecular cloud appears as a
negative source giving a high signal to noise to the central positive
part. We should clarify that the above statements are not a proof that there
is no cluster at this location. However our analysis does indicate that
even if there is a real cluster, the estimate of the SZ signal would be
biased for these sources because of the contamination even in the
NILC/MILCA maps. In particular we identify the source of contamination to
be specifically CO emission which can potentially be removed using
dedicated CO emission followup  of these sources. We also note that just as
in \cite{agha2014} we can
be less or more conservative about the cut in $\Delta \chi^2$. A less
conservative cut would permit more clusters to be identified as such at the
cost of including more contaminated sources. A more conservative approach
would result in as much of the contamination as possible to be flagged at
the risk of identifying some of the SZ signal also as contamination. We
have followed a more conservative approach in this paper.
}

\begin{table*}
\caption{\label{Tbl:Qn}Comparison of some clusters where our results differ
  from that of \cite{agha2014}. See Table \ref{Tbl:cluster} for description of columns.} 
\begin{tabular}{|c|c|c|c|c|c|c|c|}
\hline
   cluster & S/N&z&Planck valid.&$\Delta (\sum\chi^2)_{\rm
     CO-y}$ & validation &cosmology & $Q_N$\\
\hline
PSZ2 G116.05+20.00 & 5.16232 & -1.00000 & -1 & -3473.881 & MOC & F & 0.994\\
PSZ2 G124.56+25.38 & 5.35672 & -1.00000 & -1 & -883.783 & MOC & F & 1.000\\
PSZ2 G312.48-28.86 & 5.56126 & -1.00000 & -1 & -689.197 & MOC & F & 0.999\\
PSZ2 G342.32+23.51 & 4.58750 & -1.00000 & -1 & -393.434 & MOC & F & 0.998\\
PSZ2 G210.78-36.25 & 6.31805 & -1.00000 & -1 & -221.482 & MOC & F & 1.000\\
\hline
\end{tabular}
\end{table*}

The  important difference in
our algorithm when compared to \citet{agha2014} is that we also try to identify candidates which may be genuine
clusters but with significant CO contamination so that the estimate of the
$y$-distortion signal may not be accurate. The $Q_N$ of \citet{agha2014}
quantifies 
the average contamination from all sources including dust and CIB while we
focus on just one contaminant, the CO emission. Since we specifically
concentrate on the CO emission foreground, our classification can be
explicitly tested by using the ground based radio telescopes to look for
the CO emission lines with high spectral resolution. Of course neither of
the algorithms are expected to be perfect  and  our method is therefore 
complimentary to that of \citet{agha2014}. To summarize, out of a total of
130 clusters with $S/N>6$ that we identify with annotation 'MOC', 57 are
also classified by \citet{planckclusters2015} as 'Bad' with
$Q_N<0.4$. While out of 72 clusters at $S/N>6$ identified by
\citet{planckclusters2015} as 'Bad' with $Q_N<0.4$, we identify 57 as 'MOC'
and 7 as 'pMOC' and 3 as 'IND' and 4 as 'pCLG'. 

The above discussion suggests an alternative path to validating and
improving the reliability of  the Planck
cluster catalogue. This would involve using radio telescopes to measure and
subtract the CO emission in the direction of cluster candidates identified
 by us in the Planck catalogue  as $MOC$ or $pMOC$. Once an accurate CO emission has been
subtracted from HFI frequency maps at these locations, they can be combined
to make a $y$ map that is
free of CO contamination thus confirming or denying the presence of a
cluster at that location and also giving a more reliable measurement of the
$y$-distortion amplitude.
{Finally we should note that our analysis suggests that the
  recent stacking studies of $y$-distortion at positions of
  galaxies\cite{ghsb2014,rma2015} may also
  have contamination from the CO emission.}

\section{Conclusions}
We have used a parameter fitting algorithm together with model selection to separate the $y$-type distortion component from
the blackbody CMB and the foregrounds components in the Planck data. Using the same
approach we also create a CO emission map and use a second level of model
selection to identify and distinguish between $y$-type distortion and CO
emission, assuming that in any pixel one component dominates over the
other. We validate our algorithms on the real sky using known extragalactic
sources of CO emission and clusters of galaxies. The result of our analysis
is a minimal CO mask, which we make publicly available. We recommend this
mask as the minimal mask when studying $y$-type distortions in Planck
data. Our mask in particular includes regions of significant CO emission
which might be missed by masks based solely on highest frequency HFI
channels even when the masks are quite aggressive since non-negligible CO
emission is coming from the  molecular clouds and clumps where the dust
emission is relatively weak. {In addition to the $y$-maps free of
  CO contamination, our algorithm also produces CO-maps free of
  $y$-contamination and it complements the standard Dame et al. map \cite{dame2001} by
  pointing out regions of high CO emission on the sky at high galactic
latitudes not covered by the Dame et al map.}

 We revisit the Planck
catalogue and find  evidence of significant contamination from the CO
emission, particularly in the unconfirmed candidates. Our simple approach
complements the method employed by the Planck collaboration which is based on \citet{agha2014}.  We suggest a new way of validating the Planck detected clusters
by using radio telescopes to look for CO emission in the sources we have
identified as molecular clouds. These observations can then be used to
subtract the CO emission from Planck maps if non-zero emission is detected and get a clean $y$-distortion
signal if there is indeed a cluster in that direction. Alternatively if no
CO emission is detected then it would give confidence that the cluster is a
real candidate and the usual follow-up studies can be pursued. {We
  should clarify  our identification of a source as a 'MOC' just means
  that it is a potential target for radio observations from ground where
  such observation might improve the estimate and accuracy of the
  $y$-distortion signal. In particular it does not necessarily mean that there is no
  cluster present in that direction. A large observation program involving
  more than 100 scientists and targeting
  CO emission from molecular clouds identified in the Planck data will be
  underway shortly\footnote{ESO public survey Programme ID: 196.C-0999, PI:
  Ke Wang, Title: Probing the Early Stages of Star Formation: Unravelling
  the Structure of Planck Cold Clumps Distributed Throughout the Sky, \url{http://www.eso.org/sci/activities/call_for_public_surveys.html}.} \citep{kewang}.} We have
added annotations to the Planck cluster catalogue and make our annotated
catalogue also publicly available at \url{http://theory.tifr.res.in/~khatri/szresults/}.

\begin{acknowledgements}
This paper used observations obtained with Planck
(\url{http://www.esa.int/Planck}), an ESA science mission with instruments and
contributions directly funded by ESA Member States, NASA, and Canada. I
  acknowledge  use of the HEALPix software \cite{healpix}
 (\url{http://healpix.sourceforge.net}) and FFP6 simulations generated
 using  the Planck
 sky model \cite{ffp62013}
 \url{http://wiki.cosmos.esa.int/planckpla/index.php/Simulation_data}.
 This research has made use of "Aladin sky atlas" developed at CDS,
 Strasbourg Observatory, France \cite{aladin}. I also acknowledge
 useful  discussions with Eugene Churazov on component separation
 methods. I thank Rashid Sunyaev for many useful suggestions and comments
 on the manuscript.
\end{acknowledgements}

\bibliographystyle{aa}
\bibliography{co_cluster}
\end{document}